\documentclass[aps,prd,superscriptaddress,showkeys,showpacs,twocolumn,a4paper]{revtex4-1}
\usepackage{ulem}
\usepackage{color}
\usepackage{graphicx}
\usepackage{hyperref}   
\usepackage{appendix}
\usepackage{epsfig}
\usepackage{epstopdf}
\usepackage{amsmath}
\usepackage{subfig}
\usepackage{comment}
\usepackage[dvipsnames]{xcolor}
\newcommand{\be}{\begin{equation}}
\newcommand{\ee}{\end{equation}}
\newcommand{\ba}{\begin{eqnarray}}
\newcommand{\ea}{\end{eqnarray}}

\begin{document}

\title{Momentum broadening of heavy quark in a magnetized thermal QCD medium}

\author{ Balbeer Singh}
\affiliation{Theory Division, Physical Research Laboratory, Navrangpura, Ahmedabad 380 009, India}
\affiliation{Indian Institute of Technology Gandhinagar, Gandhinagar-382355, Gujarat, India}

\author{Manu Kurian}
\affiliation{Indian Institute of Technology Gandhinagar, Gandhinagar-382355, Gujarat, India}

\author{ Surasree Mazumder}
\affiliation{Theory Division, Physical Research Laboratory, Navrangpura, Ahmedabad 380 009, India}

\author{Hiranmaya Mishra}
\affiliation{Theory Division, Physical Research Laboratory, Navrangpura, Ahmedabad 380 009, India}

\author{Vinod Chandra}
\affiliation{Indian Institute of Technology Gandhinagar, Gandhinagar-382355, Gujarat, India}

\author{Santosh K. Das}
\affiliation{School of Physical Sciences, Indian Institute of Technology Goa, Ponda-403401, Goa,  India}
%\date{\today}

%%%%%%%%%%%%%%%%%%%%%%%%%%%%%%%%%%%%%%%%%%%%%%%%%%%%%%%%%%%%%%%%%%%%%%%

\begin{abstract}

Anisotropic momentum diffusion coefficients of heavy quarks have been computed in a strongly magnetized 
quark-gluon plasma beyond the static limit within the framework of Langevin dynamics. 
Depending on the orientation of the motion of the heavy quark with respect to the direction of the magnetic field, 
five momentum diffusion coefficients of heavy quark have been estimated in the magnetized thermal medium. 
Specifically, we have focussed our attention to temperature range and strength of magnetic field satisfying the condition, $\it{i.e.}$
$M\gg\sqrt{eB}\gg T$, $M$ being the mass of heavy quark. The light quarks/antiquarks follow $1+1-$dimensional lowest Landau level (LLL) kinematics, and heavy quark dynamics are not directly affected by the magnetic field in the medium. The thermal gluon contribution to the diffusion coefficient is proportions to $T^3$, whereas, the contribution of light quarks in the lowest Landau state to the same is seen to be proportional to $T|eB|$. Furthermore, it is observed that for the case of heavy quark motion parallel to the magnetic field, the component of diffusion coefficient  transverse both to the field and the heavy quark velocity $(\kappa^{\parallel}_{TT})$ turns out to be dominant as compared to the component longitudinal to both the field and motion $(\kappa^{\parallel}_{LL})$, $i.e.$, $\kappa^{\parallel}_{TT}\gg \kappa^{\parallel}_{LL}$. Further, for the case of heavy quark moving perpendicular to the magnetic field, it is seen that the diffusion coefficients transverse to the magnetic field are dominant, i.e., $\kappa^{\perp}_{LT}, \kappa^{\perp}_{TT}\gg \kappa^{\perp}_{TL}$.

%along the direction of the magnetic field and transverse to the heavy quark velocity is dominant in comparison to the components parallel and perpendicular to the heavy quark velocity in the plane transverse to the magnetic field, $i.e.$, $\kappa^{\perp}_{TL}\gg\kappa^{\perp}_{LT}\gg\kappa^{\perp}_{TT}$. Sensitivity of the various diffusion coefficients on the strength of the magnetic field and velocity of the heavy quark is also explored.

 \end{abstract}

%%%%%%%%%%%%%%%%%%%%%%%%%%%%%%%%%%%%%%%%%%%%%%%%%%%%%%%%%%%%%%%%%%%%

\keywords{Heavy quarks, Langevin dynamics, Quark-gluon plasma, Strong magnetic field}

\maketitle

%%%%%%%%%%%%%%%%%%%%%%%%%%%%%%%%%%%%%%%%%%%%%%%%%%%%%%%%%%%%%%%%%%%
 \section{Introduction}
%%%%%%%%%%%%%%%%%%%%%%%%%%%%%%%%%%%%%%%%%%%%%%%%%%%%%%%%%%%%%%%%%%% 
It is believed that intense magnetic field has been created in the initial stages of  non-central Heavy Ion Collisionsa
(HICs)~\cite{Kharzeev:NPA2008,Skokov::JMPA2009,Voronyuk:PRC2011, Deng: PRC2012,Zhong:AHEP2014,Adam:2019wnk,Acharya:2019ijj}. 
The created field is estimated to be in the order of $eB\sim m_{\pi}^2$ at Relativistic Heavy-Ion Collider (RHIC) and
a few tens of pion mass square, $eB\sim 15m_{\pi}^2$ at the Large Hadron Collider (LHC). 
Such strong magnetic fields with the strength of hadronic scale will  affect various aspects of the physics of the deconfined
hot nuclear matter created in HIC termed as Quark-Gluon Plasma (QGP).

This exciting possibility of strong magnetic field generation in HICS has led to
investigations on various characteristics of the hot nuclear matter in the presence of the magnetic field
in recent years~\cite{Karmakar:2019tdp,Dash:2020vxk,Hattori:2017qih,Kurian:2017yxj}. The study of the QGP in 
the magnetic field background opens up new avenues to explore physics in different directions such as Chiral 
Magnetic Effect (CME)~\cite{Kharzeev:NPA2008, fukushima:PRD782008}, charge-dependent elliptic 
flow~\cite{Kharzeev::PRD832011, Newman:JHEP012006, Burnier:PRL1072011, Gorbar:PRD832011}, 
magnetic catalysis of chiral symmetry breaking~\cite{Gusynin:1995nb}, various transport coefficients of QGP in the 
magnetic field~\cite{Feng:PRD962017,Fukushima:2017lvb,Kurian:2018qwb}, photon-dilepton
production~\cite{Bandyopadhyay:PRD2016, Tuchin:PRC832011,Ghosh:2018xhh}, in medium properties of
quarkonia \cite{CS:2018mag,Reddy:2017pqp} and their suppression~\cite{Hasan:EPJC2017, Singh:PRD972018}, transport coefficients 
of heavy quarks (HQs) in the magnetic field~\cite{Finazzo:2016mhm,Fukushima:2015wck, Kurian:2019nna}, etc.  

One of the challenging major uncertainties in this context is the
evolution of the magnetic field and its lifetime in hot nuclear matter. In vacuum, the magnetic field decays very rapidly. 
However, in a medium of charged particles, it can be sustained for a longer time due to induced currents arising from rapidly decreasing magnetic field.
in the medium~\cite{Tuchin:PRC2011, Fukushima:PRD2015, Mamo:PRD2015,Tuchin:PRC882013,McLerran:NPA9292014}. 
The magnetic field in the medium satisfies a diffusion equation with
diffusion coefficient $(\sigma \mu)^{-1}$, where $\mu$ is the magnetic permeability and $\sigma$ is the
electrical conductivity of the medium \cite{Tuchin:2013ie}. 
With $\mu\sim 1$ and $\sigma\simeq 0.04 T$ \cite{Yin:2013kya}, one can estimate that the time scale over which the magnetic field remains reasonably strong
over a length scale $ L$ is $\tau=L^2\sigma/4$ .
Thus, for $L\simeq 10$ fm, the time over which the magnetic field remains reasonably strong is $\tau=1$ fm for $T\sim 200$ MeV. For higher temperatures, as well as in the presence of magnetic field,  $\sigma$ will be higher, leading to  a longer time scale. However, such a conclusion is rather nontrivial because of the uncertainties associated with the non-equilibrium stage and also the expansion of the plasma has to be taken into account.  In deed, the time evolution of the magnetic field is still an open question and requires a proper estimation of electrical conductivity of the medium as well as the solutions of magneto hydrodynamics equations which needs further investigations. With these open questions in mind, it is important to explore
observables that are sensitive to the magnetic field.

To understand the properties of the QGP, one needs external probes such as highly energetic particles created at a very early stage of HICs. 
HQs serve as an effective probe to describe the properties of hot QCD medium created in the collision experiments, as they do not constitute the bulk medium, owing to their large mass compared to the temperature scale. The HQ traverses through the QGP medium as a nonthermal degree of freedom and gets random kicks from the thermal partons (light quarks/anti-quarks and gluons) in the bulk medium. Thus, the HQ dynamics could be explored within the scope of the Brownian motion~\cite{Svetitsky:1987gq,GolamMustafa:1997id,Moore:2004tg,CaronHuot:2007gq,vanHees:2005wb}, and their transport parameters, the drag and the diffusion coefficients, have been estimated in the QGP medium~\cite{Das:2013kea,Singh:2018wps,Das:2012ck,Cao:2018ews,Rapp:2018qla,Giataganas:2013zaa}. The HQ production and dynamics in the nuclear matter and the associated experimental observables have been well explored in several works~\cite{Alberico:2013bza,Dong:2019unq,Aarts:2016hap,Cao:2013ita,Song:2015sfa,Scardina:2017ipo,Adare:2006nq,Andronic:2015wma}. The HQ evolution and momentum broadening in terms of momentum diffusion in a perturbative QGP are estimated in Ref.~\cite{Moore:2004tg}. There have been some recent investigations on the estimation of the HQ momentum diffusion coefficients in a strongly magnetized QGP medium in the static limit of the HQ, i.e., HQ at rest~\cite{Fukushima:2015wck,Kurian:2019nna}. 

It turns out that in the static limit, there are two relevant diffusion coefficients of HQ in the presence of a magnetic field background, one in the direction of the magnetic field and the other perpendicular to the field. This, in turn, generates a magnetic field induced anisotropy in the diffusion coefficients.  It would be interesting to investigate the nature of the anisotropy in the diffusion of HQ beyond such static limit. The pivotal difference between the two cases in which HQ is at rest and HQ is moving is that in the later case there are two relevant directions, one is the direction of the magnetic field and, unlike the static case, the direction of the velocity of HQ. Therefore, it is imperative that the velocity of the HQ introduces another direction of anisotropy in the system. One can construct a second rank HQ diffusion tensor with five independent components, depending upon the relative orientations of the magnetic field and the velocity of HQ. 

In the current analysis, we estimate these five anisotropic diffusion coefficients of a HQ moving with finite velocity $\textbf{v}$. The HQ dynamics are described by the Langevin equations for two different cases, $viz$, the HQ moving parallel and perpendicular to the magnetic field. There can be two components of the diffusion tensor when the direction of the HQ velocity coincides with that of the magnetic field.  The second case, in which HQ moves perpendicular to the direction of the magnetic field, further introduces three components for the diffusion tensor. To estimate the HQ diffusion in the medium, we limit ourselves to the strong magnetic field and the collisions of HQ with the light partons having the soft momentum transfer so that $|\textbf{q}|\leq \sqrt{\alpha_seB}\ll T\ll \sqrt{eB}\ll M$, where $|\textbf{q}|$ is the magnitude of the momentum transfer from the HQ to light thermal partons. For this purpose, we shall use the resummed gluon propagator at finite temperature and magnetic field. In a recent work~\cite{Singh:2020fsj}, some of the authors of the present paper
discussed the collisional energy loss of the HQ using a similar technique, which might throw some insights into the jet quenching. On the other hand, in the present calculation the momentum broadening of HQ in various directions depending upon the relative orientation of the magnetic field and the velocity of the HQ have been estimated. We follow an approach similar to the Refs.~\cite{Moore:2004tg, Romatschke:2006bb} where effects of the magnetic field are not considered. 

We found the ratio, $\kappa^{\parallel}_{LL}/\kappa^{\parallel}_{TT}\ll 1$ which is similar to the result found in Ref.~\cite{Fukushima:2015wck}. Our notation for the diffusion coefficients are as follows. The superscript denotes the HQ velocity with respect to the magnetic field. Of the two subscripts, the first one describes the momentum diffusion relative to the direction of the velocity of the HQ while the second index refer the momentum diffusion relative to the direction to the magnetic field. Further, for the case in which HQ moves perpendicular to the magnetic field, the diffusion coefficients in the plane transverse to the magnetic field are larger than that in the direction of the magnetic field, $\kappa^{\perp}_{LT},\kappa^{\perp}_{TT}\gg \kappa^{\perp}_{TL}$. The velocity dependence of the quark contribution of these diffusion components are very much similar to those found in the Ref.~\cite{Mamo:2016prd} in the context of jet quenching.  

The manuscript is organized as follows. In section~\ref{formalism}, we discuss the Langevin formalism for HQ diffusion for both the cases, $i.e.$, HQ moving parallel and perpendicular to the magnetic field. In subsections~\ref{gluonic} and~\ref{fermionic}, we estimate the contribution to the diffusion coefficients from the gluon and the light quarks/anti-quarks, respectively where the light quark mass has been neglected. We have given a brief description of the scenario with non-vanishing light quark mass in section~{\ref{non zero quark mass}}.
In section~{\ref{results}}, we discuss in detail the results of the present investigation. Finally, in section~{\ref{summary}}, we summarise the present work and discuss the implications and future possibilities. In Appendix~\ref{selfenergy}, we discuss the details of the calculation of gluon self energy in the magnetic field in the LLL approximation. 

%%%%%%%%%%%%%%%%%%%%%%%%%%%%%%%%%%%%%%%%%%%%%%%%%
{\bf{Notations}}:
\label{setup}
The magnetic field is considered to be constant and along $\textbf{z}$-axis so that ${\bf B}=B\hat{z}$.
The calculations of the quark propagator in Real-Time formalism and the relevant matrix element squared require the following notations, where $\parallel$ and $\perp$ represent the components parallel and perpendicular to the magnetic field, respectively. For the metric tensor, we use
\begin{equation}
g_{\mu \nu}^{\parallel}=(1,0,0,-1),  \hspace{1cm} g_{\mu \nu}^{\perp}=(0,-1,-1,0).
\end{equation}
The parallel ($i.e.$, $a_{\mu}^{\parallel}=g_{\mu \nu}^{\parallel}a^{\nu}$) and perpendicular ($i.e.$, $a_{\mu}^{\perp}=g_{\mu \nu}^{\perp}a^{\nu}$) components of a four-vector $a_{\mu}$ are represented as
\begin{equation}
a_{\mu}^{\parallel}=(a_{0},0,0,-a_{3}), \hspace{1cm} a_{\mu}^{\perp}=(0,-a_{1},-a_{2},0). 
\end{equation}
The four-vector product ($a^{\mu}b_{\mu}=a\cdot b$) can be written as
\begin{equation}
a \cdot b=a_{\parallel}\cdot b_{\parallel}-a_{\perp}\cdot b_{\perp}.
\end{equation}
Similarly, the square of both the components of the four-vector can be denoted as,
\begin{equation}
a^2_{\parallel}=a_{0}^2-a_{3}^2,  \hspace{1cm} a^2_{\perp}=a_{1}^2+a_{2}^2.
\end{equation}
For four momentum vector we use the notation $K_{\mu}=(k_0,-\textbf{k})$ with the parallel component $k^{\parallel}_{\mu}=(k_0,0,0,-k_z)$ and the perpendicular component $k^{\perp}_{\mu}=(0,-k_x,-k_y,0)$. 

%%%%%%%%%%%%%%%%%%%%%%%%%%%%%%%%%%%%%%%%%%%%%%%%%%%%%%%%%%%%%%%%%%%%
\section{Formalism: Langevin dynamics of heavy quark in a magnetized medium} 
\label{formalism}
%%%%%%%%%%%%%%%%%%%%%%%%%%%%%%%%%%%%%%%%%%%%%%%%%%%%%%%%%%%%%%%%%%%%
We will work in the strong magnetic field limit with $\sqrt{eB}\gg T$, indicating that the light quarks/antiquarks occupy the lowest Landau level (LLL) while thermal gluons are unaffected by the field. Note that the HQ motion is not Landau quantized as $M\gg \sqrt{eB}$. 
To estimate the thermal gluons and thermal light quark/antiquark contributions to the transport coefficients of the HQ for the non-static case, $i.e.$, when the HQ is moving with velocity $\bf{v}$ in the medium, we consider two cases: when HQ is moving along the direction of the magnetic field ($\bf v \parallel \bf B$) and when the HQ motion is transverse to the magnetic field ($\bf v\perp \bf B$). 

%%%%%%%%%%%%%%%%%%%%%%%%%%%%%%%%%%%%%%%%%%%%%%%%%%%%%%%%%%%%%%%%%%%%
\subsubsection*{Case I: {\bf v} ${\parallel}$ {\bf B}}
\label{case1}
%%%%%%%%%%%%%%%%%%%%%%%%%%%%%%%%%%%%%%%%%%%%%%%%%%%%%%%%%%%%%%%%%%%%

\begin{figure}[h]
    \includegraphics[width=0.225\textwidth]{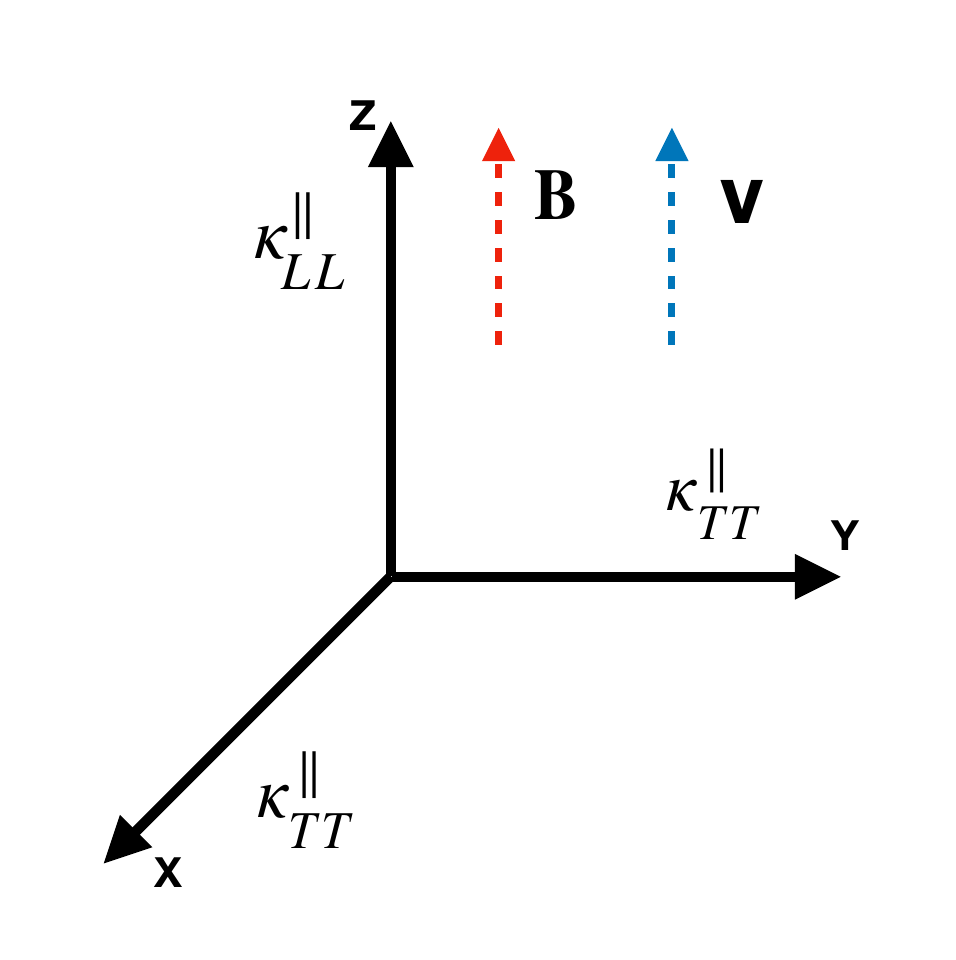}
    \caption{HQ motion parallel to magnetic field}
    \label{fig1}
\end{figure}

The magnetic field $\bf B$, and the HQ velocity $\bf v$, are considered to be in the same direction as depicted in Fig.~\ref{fig1}. 
The general structure of HQ momentum diffusion tensor in this case can be decomposed as follows,
\begin{align}\label{1}
 \kappa^{ij}=R^{ij}\kappa^{\parallel}_{TT}+Q^{ij}\kappa^{\parallel}_{LL},
\end{align}
where $R^{ij}=\big(\delta^{ij}-\frac{p^ip^j}{p^2}\big)$ and $Q^{ij}=\frac{p^ip^j}{p^2}$ are the transverse and longitudinal projection operators orthogonal to each other, $i.e.$, $R^{ij}Q_{ij}=0$. Here, $\kappa^{\parallel}_{TT}$ and $\kappa^{\parallel}_{LL}$ are the two diffusion coefficients, transverse and longitudinal to the direction of HQ motion (which is same as the direction of $\bf B$). The symbol $\parallel$ denotes that the HQ motion is parallel to the direction of the magnetic field.
The broadening of the variance of HQ momentum distribution can be described by the macroscopic equation of motion as follows~\cite{Moore:2004tg},
\begin{align}\label{2}
&\dfrac{d}{dt}\langle p\rangle=-\eta^{\parallel}_{D}(p)p,\nonumber\\
&\dfrac{1}{2}\dfrac{d}{dt}\langle(\Delta p_T)^2\rangle=\kappa^{\parallel}_{TT}(p),\nonumber\\
&\dfrac{d}{dt}\langle(\Delta p_L)^2\rangle=\kappa^{\parallel}_{LL}(p),
\end{align}
where the coefficient $\eta^{\parallel}_D$ measures the average momentum loss. The variance of the HQ momentum distribution transverse and parallel to the direction of the motion can be respectively defined as $\langle(\Delta p_T)^2\rangle\equiv\langle p_T^2\rangle$ and $\langle(\Delta p_L)^2\rangle\equiv (p_L-\langle p_L\rangle)^2$. The factor $\frac{1}{2}$ in the transverse momentum broadening is
due to two perpendicular directions. The HQ transport coefficients $\eta^{\parallel}_D$, $\kappa^{\parallel}_{LL}$ and $\kappa^{\parallel}_{TT}$ can be obtained from the kinetic theory by considering the proper collisional scattering amplitude and have the following form for the rate of average momentum loss,
\begin{align}
\dfrac{d}{dt}\langle p\rangle&=\dfrac{1}{2v}\int_{k, q}
|\bar{\mathcal{M}}|^2 \omega \bigg[f(k)\Big(1\pm f(k+\omega)\Big)\nonumber\\
&-f(k+\omega)\Big(1\pm f(k)\Big)\bigg],   
\end{align}
where $v=|\textbf{p}|/E$ is the magnitude of the velocity of HQ, and $f$ is the distribution of thermal particles in the magnetized QGP, $|\bar{\mathcal{M}}|$ is the HQ-thermal particle scattering matrix element, and $\omega$ is the transferred energy due to the scattering process of the HQ with the medium partons. We can write similar expressions for the rate of transverse and longitudinal momentum broadening which are $\kappa^{\parallel}_{TT}(\bf v)$ and $\kappa^{\parallel}_{LL}(\bf v)$, $i.e.$,
\begin{align}
\kappa^{\parallel}_{TT}(\bf{v})&=\int_{k, q}{|\bar{\mathcal{M}}|^2}q_T^2\bigg[f(k)\Big(1\pm f(k+\omega)\Big)\bigg],\\ 
\kappa^{\parallel}_{LL}(\bf v)&=\int_{k, q}{|\bar{\mathcal{M}}|^2}q_z^2\bigg[f(k)\Big(1\pm f(k+\omega)\Big)\bigg].\nonumber\\
\end{align}
where $q_T$ and $q_z$ are the magnitude of the transverse and longitudinal momentum transfer in the scattering process. The notation $\int_{k,q}$ denotes the relevant phase space integration over $\bf k$ and $\bf q$ with proper dimensions. In the case of small energy transfer, i.e., $\omega=\textbf{v}\cdot \textbf{q}$, one can write
\begin{equation}
f(k)(1\pm f(k+\omega))-f(k+\omega)(1\pm f(k))\approx \frac{\omega}{T}f(k)(1\pm f(k) ),
\end{equation}
and
\begin{equation}
f(k)(1\pm f(k+\omega))\approx f(k)(1\pm f(k)).
\end{equation}
We shall use these approximation in evaluating the diffusion coefficients. 
%%%%%%%%%%%%%%%%%%%%%%%%%%%%%%%%%%%%%%%%%%%%%%%%%%%%%%%%%%%%%%%%%%%%
\subsubsection*{Case II: {\bf v} $\perp$ {\bf B}} 
\label{case2}
%%%%%%%%%%%%%%%%%%%%%%%%%%%%%%%%%%%%%%%%%%%%%%%%%%%%%%%%%%%%%%%%%%%%%

When the HQ is moving transverse to the direction of the magnetic field, say, ${\bf{v}}=(v_x, 0, 0)$ and ${\bf{B}}=B\hat{z}$, the momentum broadening can be characterized by three diffusion coefficients.
\begin{figure}[h]
\includegraphics[width=0.225\textwidth]{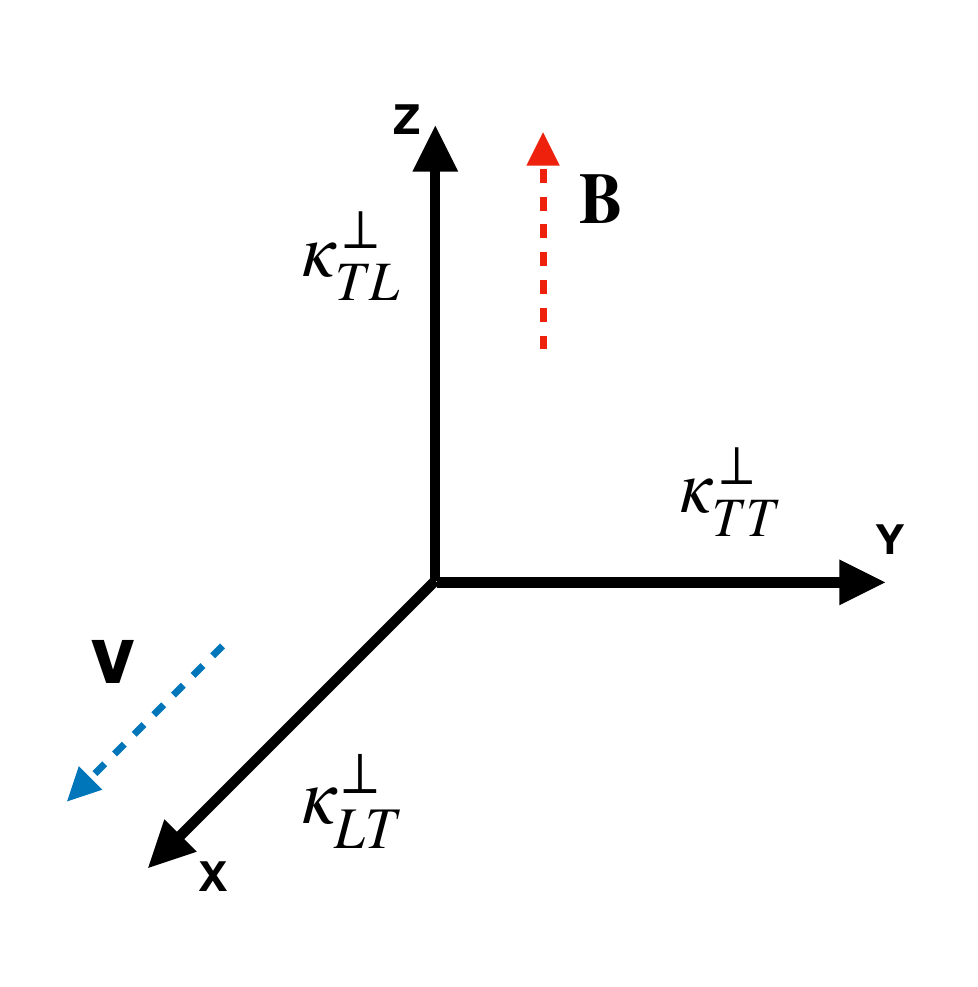}
\caption{HQ is moving in the $\textbf{x}$-axis in the presence of the strong magnetic field ${\bf{B}}=B\hat{z}$.}
\label{fig2}
\end{figure}
Defining ${\bf{b}}=(0, 0, 1)$ to project the direction of magnetic field, the diffusion tensor can be decomposed as follows,
\begin{align}\label{1}
   \kappa^{ij}=P^{ij}\kappa^{\perp}_{TL}+Q^{ij}\kappa^{\perp}_{LT}+R^{ij}\kappa^{\perp}_{TT},
\end{align}
in which the projection operators takes the forms,
\begin{align}
&P^{ij}=\frac{b^ib^j}{b^2}, &Q^{ij}=\frac{p^ip^j}{p^2}, &&R^{ij}=\bigg(\delta^{ij}-\frac{p^ip^j}{p^2}-\frac{b^ib^j}{b^2}\bigg),   
\end{align}
such that the operators are orthogonal to each other. Note that the same decomposition is valid for the HQ motion along $\textbf{y}$-axis. In this case, the Langevin equations take the forms as,
\begin{align}
&\dfrac{d}{dt}\langle p\rangle=-\eta^{\perp}_{D}(p)p,
&&\dfrac{d}{dt}\langle(\Delta p_z)^2\rangle=\kappa^{\perp}_{TL}(p),\nonumber\\
&\dfrac{d}{dt}\langle(\Delta p_x)^2\rangle=\kappa^{\perp}_{LT}(p),
&&\dfrac{d}{dt}\langle(\Delta p_y)^2\rangle=\kappa^{\perp}_{TT}(p).
\end{align}
The component $\kappa^{\perp}_{TL}$ denotes the diffusion coefficient in the direction transverse to the HQ motion and longitudinal to the direction of the magnetic field, $i.e.$, along $\textbf{z}$-axis. Similarly, $\kappa^{\perp}_{LT}$ and $\kappa^{\perp}_{TT}$ respectively define the components of the diffusion coefficient longitudinal to the HQ motion and transverse to the magnetic field (along $\textbf{x}$-axis), and in the direction transverse to both HQ motion and magnetic field (along $\textbf{y}$-axis). The diffusion coefficients are estimated from the following expressions~\cite{Moore:2004tg,Romatschke:2006bb},
\begin{align}
\kappa^{\perp}_{TL}(\bf v)&=\int_{k, q}{|\bar{\mathcal{M}}|^2}q_z^2\bigg[f(k)\Big(1\pm f(k+\omega)\Big)\bigg],\\ 
\kappa^{\perp}_{LT}(\bf v)&=\int_{k, q}{|\bar{\mathcal{M}}|^2}q_x^2\bigg[f(k)\Big(1\pm f(k+\omega)\Big)\bigg],\\
\kappa^{\perp}_{TT}(\bf v)&=\int_{k, q}{|\bar{\mathcal{M}}|^2}q_y^2\bigg[f(k)\Big(1\pm f(k+\omega)\Big)\bigg].\nonumber\\
\end{align}
In the following section, we discuss the interaction of HQ with the thermal gluon and the light quark/antiquark in detail. 

\section{Diffusion in the massless light quark limit}
In this section, we discuss the HQ diffusion coefficient in the limit when the light quark mass is zero, for both $\textbf{v}\parallel\textbf{B}$ and $\textbf{v}\perp\textbf{B}$ cases. Moreover, in this limit, the thermal contribution to the self energy vanishes; hence, the diffusion coefficients get the contribution from the magnetic field dependent self energy only. However, on the other hand, for finite light quark mass, both thermal as well as non-thermal part of the self energy give finite contributions to the diffusion coefficients.
 
%%%%%%%%%%%%%%%%%%%%%%%%%%%%%%%%%%%%%%%%%%%%%%%%%%%%%%%%%%%%%%%%%%%%%
\subsection{Gluonic contribution} 
\label{gluonic}
%%%%%%%%%%%%%%%%%%%%%%%%%%%%%%%%%%%%%%%%%%%%%%%%%%%%%%%%%%%%%%%%%%%%%
Gluonic contribution to the diffusion coefficient comes via the Compton scattering, $i.e.$, $Q(P)+g(K)\rightarrow Q(P^{'})+g(K^{'})$, 
where $g$ stands for gluon. Generally, at leading order in the coupling, there are three channels, $s, t$ and $u$ that contribute to 
Compton scattering. In the small momentum transter limit, the magnetic field contribution arises from the $t$-channel scattering. 
This is because the contribution from the $s$ and the $u$-channels of the Compton scattering is negligible in the presence of magnetic 
field due the hierarchy in the scales considered here, $i.e.$, in the regime $M\gg\sqrt{eB}$. Therefore, the HQ propagators appearing in $s$ and $u$ channels 
are not affected by the magnetic field. In the $t$-channel scattering, the effect of the magnetic field comes through the resummed gluon 
propagator. For a HQ at rest, the matrix elements for the $t$-channel scattering  is well investigated 
in the Ref.~\cite{Fukushima:2015wck} using the Debye mass screened gluon propagator. In contrast, when a HQ is not static and is having a finite velocity, we use the resummed gluon propagator for the low momentum transfer processes as relevant for the estimation of diffusion coefficients. Accordingly, the color-averaged $t$-channel scattering amplitude takes the form~\cite{Singh:2020fsj}, 
\begin{eqnarray}
|\mathcal{\bar{M}}|^2&=&\frac{64 g^4(q_{\parallel}^2\Pi^{\parallel}_R)^2 }{Q^4(Q^2-q_{\parallel}^2\Pi^{\parallel}_R)^2}\bigg(\mathcal{A}-M^2(K.P_{\parallel}.K')\bigg)\nonumber\\
&-&\frac{64g^4q_{\parallel}^2\Pi^{\parallel}_R}{Q^4(Q^2-q_{\parallel}^2\Pi^{\parallel}_R)}\bigg(\mathcal{B}+2M^2(K.P_{\parallel}.K')\bigg)\nonumber\\
&+&\frac{64g^4}{Q^4}\bigg((P.K)(P'.K')+(P.K')(P'.K)\bigg),
\label{12}
\end{eqnarray}
where,
\begin{eqnarray}
\mathcal{A}&=&(P.P_{\parallel}.K)(P'.P_{\parallel}.K')
+(P.P_{\parallel}.K')(K.P_{\parallel}.P')\nonumber\\
&+&(P.P')(K.P_{\parallel}.K'),
\end{eqnarray}
\begin{eqnarray}
\mathcal{B}&=&(P.K)(P'.P_{\parallel}.K')+(P.K')(K.P_{\parallel}.P')+(K.P')\nonumber\\
&\times&(P'.K')+(P.P_{\parallel}.K)-2(P.P')(K.P_{\parallel}.K').
\end{eqnarray}
Here, 
\begin{eqnarray}
P.P_{\parallel}.K=& P_{\mu}P_{\parallel}^{\mu \nu}K_{\nu}=&\frac{(P.q_{\parallel})(K.q_{\parallel})}{q_{\parallel}^2}-P.k_{\parallel},
\end{eqnarray}
where $Q=(\omega,\textbf{q})=K'-K=P-P'$, is the four momentum vector for the exchange gluon and $q_{\parallel}^2=\omega^2-q_z^2$. 
%In the above equation
%\begin{equation}
%P^{\mu \nu}_{\parallel}=-g^{\mu \nu}_{\parallel}+\frac{q^{\mu}_{\parallel} q^{\nu}_{\parallel}}{q_{\parallel}^2}
%\end{equation}
%which is the longitudinal projection operator. 
For the estimation of the diffusion and the drag coefficients, we restrict the energy transfer to be small, which can be done by approximating $\omega=\textbf{v}\cdot \textbf{q}$ which arises from the energy conservation. Here $\textbf{q}$ is the momentum transfer between HQ and thermal partons. Further, the gluon retarded self energy,$\Pi^{\parallel}_{R}$, in Eq.\ref{12}, is explicitely given in Appendix~\ref{selfenergy}. It is evident from Eqs.~(\ref{self3}) and (\ref{self2}) that $\Pi^{\parallel}_R$ at zero temperature has a part proportional to the square of the light quark mass.On the other hand, as shown in Eq.\ref{pir1},  the thermal part of $\Pi^{\parallel}_R$ is proportional to the light quark mass. We will first consider the case where the light quarks are assumed to be massless and discuss the finite mass effects later.

\subsubsection*{Case I: {\bf v} $\parallel$ {\bf B}}

Let us first discuss the case in which the HQ is moving along the direction of the magnetic field, i.e., along the z-axis(as shown in Fig.~(\ref{fig1})). One can write down the specific expression for the square of the amplitude in this
particular case in the following concise form: 
\begin{align}
|\mathcal{\bar{M}}|^2&=\frac{64g^4}{(Q^2-q_{\parallel}^2\Pi^{\parallel}_R)^2}\bigg[2(P.K)^2-2(P.K)(K.Q)\nonumber\\
& -Q^2(P.K)+ 2\frac{q_{\parallel}^2\Pi^{\parallel}_R}{Q^2}(P.K)(K.Q)+q_{\parallel}^2\Pi^{\parallel}_R (P.K)\bigg].
\label{amplparallel}
\end{align}
The present calculation is performed in the small momentum transfer limit, $\it{i.e.}$ when $|\textbf{q}|^2\sim \alpha_s eB$ and according to the hierarchy of scales 
$T^2 \gg \alpha_s eB$. Therefore, the most dominant terms will be those proportional to $\log(T^2/\alpha_s eB)$. After
simplifying the expressions of the two diffusion coefficients, it is 
possible to identify the terms from Eq.~(\ref{amplparallel}) contributing to the leading-log; $\it{i.e.}$ $\log(T^2/\alpha_s eB)$ order.
The gluonic contribution to diffusion coefficient along the direction of the magnetic field takes the form~\cite{Moore:2004tg},
\begin{eqnarray}
\kappa^{\parallel}_{LL;Qg}&=&\frac{1}{16 E^2}\int \frac{d\textbf{k}}{(2\pi)^3|\textbf{k}|}\frac{d\textbf{k}'}{(2\pi)^3|\textbf{k}'|}
\frac{d\textbf{p}'}{(2\pi)^3}q_z^2|\mathcal{\bar{M}}|^2f(|\textbf{k}|)\nonumber\\
&\times&(1+f(|\textbf{k}'|))(2\pi)^4\delta^{4}(P+K-P'-K'),
\label{15}
\end{eqnarray} 
where $f(k)$ is Bose-Einstein distribution function and $|\textbf{k}|=k$ and $|\textbf{k}'|=k'$.
After performing $\textbf{p}'$ integration by using the three momenta Dirac delta function, Eq.~(\ref{15}) reduces to
\begin{eqnarray}
\kappa^{\parallel}_{LL;Qg}&=&\frac{1}{16 E^2(2\pi)^5}\int \frac{d\textbf{k}}{k}\frac{d\textbf{k}'}{k'}q_z^2|\mathcal{\bar{M}}|^2f(k)\nonumber\\
&\times&(1+f(k'))\delta(E+k-E'-k').
\label{16}
\end{eqnarray}
Introducing the identity $\int d\textbf{q}~\delta^3({\bf k'}-{\bf k}-{\bf q})=1$, one can arrive at
\begin{eqnarray}
\kappa^{\parallel}_{LL;Qg}&=&\frac{1}{16 E^2(2\pi)^5}\int \frac{d\textbf{k}}{k}\frac{d\textbf{q}}{k'}q_z^2|\mathcal{\bar{M}}|^2f(k)\nonumber\\
&\times&(1+f(k'))\delta(E+k-E'-k').
\label{17}
\end{eqnarray}
In the small momentum transfer limit.
%The Iase in which the momentum transfer is small, one may use $k'\approx k$ in the denominator and $f(k)(1+f(k'))\approx f(k)(1+f(k))$.
using the approximations $E-E'\approx \textbf{v}.\textbf{q}$ and $k'-k\approx \hat{k}.\textbf{q}$, the diffusion coefficient becomes
\begin{eqnarray}
\kappa^{\parallel}_{LL;Qg}&=&\frac{1}{16 E^2(2\pi)^5}\int \frac{d\textbf{k}}{k^2}~d\textbf{q}q_z^2|\mathcal{\bar{M}}|^2f(k)\nonumber\\
&\times&(1+f(k))\delta(\textbf{v}.\textbf{q}-\hat{k}.\textbf{q}).
\label{18}
\end{eqnarray}
At this point, we use such a co-ordinate system in which the vectors, $\textbf{k}$ and $\textbf{q}$ make angles $\theta_k$ and $\theta_q$
with z-axis(and with HQ velocity in this case) and azimuthal angles $\phi_k$ and $\phi_q$, respectively. One may perform the $\phi_k$-integration
with the help of the delta function left inside the integral by rearranging the delta function as
\begin{equation}
\delta(\textbf{v}.\textbf{q}-\hat{k}.\textbf{q})=\delta(f(\phi_k))=\frac{\delta(\phi_k-\phi_k^0)}{|f'(\phi_k^0)|},
\label{19}
\end{equation}
where, $\phi_k^0$ is the root of the function $f(\phi_k)=0$ and described by
\begin{equation}
\phi_k^0=\phi_q+\cos^{-1}\bigg(\frac{v\cos\theta_q-\cos\theta_k \cos\theta_q}{\sin\theta_k \sin\theta_q}\bigg).
\label{20}
\end{equation}
With 
\begin{equation}
|f'(\phi_k^0)|=q|\sin\theta_k||\sin\theta_q||\sin(\phi_k^0-\phi_q)|,
\label{21}
\end{equation}
the diffusion coefficient becomes
\begin{eqnarray}
\kappa^{\parallel}_{LL;Qg}&=&\frac{1}{16 E^2(2\pi)^5}\int^{1}_{-1} \frac{\cos^2\theta_q}{|\sin\theta_q|}d(\cos\theta_q)\nonumber\\
&\times&\int_{0}^{2\pi}\frac{d\phi_q}{|\sin(\phi_k^0-\phi_q)|}
\int \frac{d(\cos\theta_k)}{|\sin\theta_k|}\nonumber\\
&\times& \int^{\infty}_{0}f(k)(1+f(k))dk\int_0^{2k}q^3dq|\mathcal{\bar{M}}|^2.
\label{22}
\end{eqnarray}
The limits of the $\cos\theta_k$-integration is fixed by the condition $|\cos(\phi_k^0-\phi_q)|\leq 1$ and that of q-integration 
by the expression $\cos\theta_{kk'}=1-q^2/2k^2$ in small momentum transfer limit. It is evident from Eq.~(\ref{22}) that the terms in
$|\mathcal{\bar{M}}|^2$ contributing to $\log(T^2/\alpha_s eB)$ order arise from the first and the last term of Eq.\ref{amplparallel}. The other terms will give sub dominant contribution which can clearly be observed from Eq.~(\ref{22}).We will represent the expression for both the cases. 

%\subsubsection*{Case I: {\bf v} $\parallel$ {\bf B}} 
The first term of Eq.\ref{amplparallel}, contributing to the leading-log order, can be expressed in the current frame of reference as
\begin{equation}
|\mathcal{\bar{M}}|^2=\frac{128g^4}{(Q^2+m_{D,B}^2)^2}E^2k^2(1-v\cos\theta_k)^2,
\end{equation}
where $m_{D,B}$ is the \textquotedblleft effective" Debye mass given by
\begin{equation}
m_{D,B}^2=-q_{\parallel}^2\Pi^{\parallel}_R
\label{debyemass}
\end{equation}
The part of the above expression which contributes to the leading logarithmic or $\log(T^2/\alpha_s eB)$ order in the diffusion coefficient
is 
\begin{eqnarray}
\kappa^{\parallel}_{LL;Qg}&=& \frac{4T^3g^4\zeta(2)}{(2\pi)^4}\log\bigg(\frac{2T}{m_{D,B}}\bigg)\int_{-1}^{1}\frac{z^2dz}{(v^2z^2-1)^2}\nonumber\\
&\times& \int \frac{(1-vt)^2 dt}{\sqrt{1-z^2-t^2-v^2z^2+2vtz^2}},\nonumber\\
&\equiv& \frac{4T^3g^4\zeta(2)}{(2\pi)^4}\log\bigg(\frac{2T}{m_{D,B}}\bigg)\mathcal{E}(v),
\label{23}
\end{eqnarray}
where $z=\cos\theta_q$ and $t=\cos\theta_k$.

Similarly, the logarithmic contribution of the other component of the diffusion coefficient orthogonal to the magnetic field as well 
as HQ velocity can be written as
\begin{eqnarray}
\kappa^{\parallel}_{TT;Qg}&=& \frac{4T^3g^4\zeta(2)}{(2\pi)^4}\log\bigg(\frac{2T}{m_{D,B}}\bigg)\int_{-1}^{1}\frac{(1-z^2)dz}{(v^2z^2-1)^2}\nonumber\\
&\times& \int \frac{(1-vt)^2 dt}{\sqrt{1-z^2-t^2-v^2z^2+2vtz^2}},\nonumber\\
&\equiv& \frac{4T^3g^4\zeta(2)}{(2\pi)^4}\log\bigg(\frac{2T}{m_{D,B}}\bigg)\mathcal{F}(v).
\label{24}
\end{eqnarray}
where $\mathcal{E}(v)$ and $\mathcal{F}(v)$ are results of the angular integrations which are functions of the HQ velocity. It is instructive to mention that Eq.\ref{23} is found to reduce to the result obtained in Ref.~\cite{Fukushima:2015wck} in the limit when the HQ velocity is zero.

From Eq.~(\ref{amplparallel}), one can see that there is another term, $q_{\parallel}^2\Pi^{\parallel}_R(P.K)$ or $m_{D,B}^2(P.K)$, which will have a leading log contribution.
The expressions for the two diffusion coefficients for this particular term are
\begin{eqnarray}
\kappa^{\parallel}_{LL;Qg}&=& \frac{T^2g^4m_{D,B}^2\zeta(1)}{2E(2\pi)^4}\log\bigg(\frac{2T}{m_{D,B}}\bigg)\mathcal{G}(v),\nonumber\\
\kappa^{\parallel}_{TT;Qg}&=& \frac{T^2g^4m_{D,B}^2\zeta(1)}{2E(2\pi)^4}\log\bigg(\frac{2T}{m_{D,B}}\bigg)\mathcal{H}(v),
\label{25}
\end{eqnarray}
where, 
\begin{eqnarray}
\mathcal{G}(v)&=& \int^{1}_{-1}\frac{z^2dz}{(v^2z^2-1)^2}\nonumber\\
&\times& \int \frac{(1-vt)dt}{\sqrt{1-t^2-z^2-v^2z^2+2vtz^2}},\nonumber\\
\mathcal{H}(v)&=& \int^{1}_{-1}\frac{(1-z^2)dz}{(v^2z^2-1)^2}\nonumber\\
&\times& \int \frac{(1-vt)dt}{\sqrt{1-t^2-z^2-v^2z^2+2vtz^2}}.
\label{26}
\end{eqnarray}
It is evident from Eq.~(\ref{25}) that the contribution due to the term proportional to $q_{\parallel}^2\Pi^{\parallel}_R(P.K)$ are of the order $g^6$, whereas
the contribution due to the term proportional to $(P.K)^2$ are of the lower order in g, $\it{i.e.}$, $g^4$. Therefore, when $g\ll 1$, one can drop the terms which are of the order $g^6$ in comparison to the term of the $g^4$ order. Consequently, one can argue that for small couplings, i.e., $g\ll 1$, the most dominant terms for both the diffusion coefficients are given by Eq.~(\ref{23}) and Eq.~(\ref{24}).

\subsubsection*{Case II: {\bf v} $\perp$ {\bf B}}
Now, let us consider the other case where HQ moves perpendicular to the magnetic field, as shown in Fig.~(\ref{fig2}). 
Without losing generality, we choose the HQ motion along the ${\textbf{x}}$-axis so that the HQ velocity takes the 
form $\textbf{v}=(v\hat{\textbf{x}},0,0)$. As mentioned earlier, in this case, there are three diffusion coefficients, 
$\kappa^{\perp}_{LT;Qg}, \kappa^{\perp}_{TL;Qg}$, and $\kappa^{\perp}_{TT;Qg}$. just like the previous case, it is our aim
in this case, too, to write down the leading log terms. From Eq.~(\ref{12}), after
some trivial algebra, one can find out those terms of $|\mathcal{\bar{M}}|^2$ which will produce results in leading-log:
\begin{eqnarray}
|\mathcal{\bar{M}}|^2&=& \frac{64g^4}{(Q^2-q_{\parallel}^2\Pi^{\parallel}_R)^2}\bigg(\frac{q_{\parallel}^2\Pi^{\parallel}_R}{Q^2}E\omega (K.Q)+2q_{\parallel}^2\Pi^{\parallel}_R(P.K)\nonumber\\
&+&\Pi^{\parallel}_R
Ekq_z(q_z-\omega\cos\theta_k)+2(P.K)^2\nonumber\\
&+& 2\frac{q_{\parallel}^2\Pi^{\parallel}_R}{Q^2}(K.Q)(P.Q)\bigg).
\label{27}
\end{eqnarray}
The integrations for the three diffusion coefficients are performed in the similar way, assuming a similar co-ordinate system same as $\it{Case:I}$ having
the HQ velocity along the x-axis. Therefore, the diffusion coefficient along the HQ velocity and perpendicular to the 
magnetic field can be written as
\begin{eqnarray}
\kappa^{\perp}_{LT;Qg}&=& \frac{1}{16E^2(2\pi)^5}\int^{1}_{-1}\frac{d(\cos\theta_q)}{|\sin\theta_q|}\int \frac{d(\cos\theta_k)}{|\sin\theta_k|}\nonumber\\
&\times& \int^{2\pi}_{0}\frac{d\phi_q}{|\sin(\phi_k^0-\phi_q)|}
\int^{\infty}_{0}
f((k)(1+f(k))dk \nonumber\\
&\times& \int^{2k}_{0}q dq q_x^2|\mathcal{\bar{M}}|^2,\nonumber\\
&=& \frac{1}{16E^2(2\pi)^5}\int^{1}_{-1}\frac{\sin^2\theta_q d(\cos\theta_q)}{|\sin\theta_q|}\int \frac{d(\cos\theta_k)}{|\sin\theta_k|}\nonumber\\
&\times& \int^{2\pi}_{0}\frac{\cos^2\phi_q d\phi_q}{|\sin(\phi_k^0-\phi_q)|}\int^{\infty}_{0}f((k)(1+f(k))dk\nonumber\\
&\times& \int^{2k}_{0}q^3 dq |\mathcal{\bar{M}}|^2,
\label{28}
\end{eqnarray}
with a different $\phi_k^0$ than that in the previous case of $\textbf{v}\parallel \textbf{B}$:
\begin{equation}
\phi_k^0= \phi_q+\cos^{-1}\bigg(\frac{v\sin\theta_q\cos\phi_q-\cos\theta_k\cos\theta_q}{\sin\theta_k\sin\theta_q}\bigg).
\label{29}
\end{equation}
It is easy to identify, from Eq.~(\ref{27}), that only the fourth term in the brackets in Eq.~\ref{27}, $\it{i.e.}$ the term $2(P.K)^2$, will contribute to the leading log and $g^4$. The other terms, though will be logarithmic, will be of $g^6$ order which one can neglect in comparison to the $g^4$ terms for small couplings.
Therefore, the most dominant terms of the three diffusion coefficients are:
\begin{eqnarray}
\kappa^{\perp}_{LT;Qg}&=& \frac{4T^3g^4\zeta(2)}{(2\pi)^5}\log\bigg(\frac{2T}{m_{D,B}}\bigg)\mathcal{J}(v),\nonumber\\
\kappa^{\perp}_{TL;Qg}&=& \frac{4T^3g^4\zeta(2)}{(2\pi)^5}\log\bigg(\frac{2T}{m_{D,B}}\bigg)\mathcal{K}(v),\nonumber\\
\kappa^{\perp}_{TT;Qg}&=& \frac{4T^3g^4\zeta(2)}{(2\pi)^5}\log\bigg(\frac{2T}{m_{D,B}}\bigg)\mathcal{L}(v).
\label{30}
\end{eqnarray}
where $\mathcal{J}(v)$, $\mathcal{K}(v)$ and $\mathcal{L}(v)$ are the results of the angular integrations and 
functions of HQ velocity. Therefore,
\begin{eqnarray}
\mathcal{J}(v)&=& \int^{1}_{-1}\frac{\sin^2\theta_q d(\cos\theta_q)}{|\sin\theta_q|}\int^{2\pi}_{0}\frac{\cos^2\phi_q d\phi_q}{(v^2\sin^2\theta_q\cos^2\phi_q-1)^2}\nonumber\\
&\times& \int \frac{(1-v\sin\theta_k\cos\phi_k^0)^2 d(\cos\theta_k)}{|\sin\theta_k||\sin(\phi_k^0-\phi_q)|},
\label{31}
\end{eqnarray}
\begin{eqnarray}
\mathcal{K}(v)&=& \int^{1}_{-1}\frac{\cos^2\theta_q d(\cos\theta_q)}{|\sin\theta_q|}\int^{2\pi}_{0}\frac{d\phi_q}{(v^2\sin^2\theta_q\cos^2\phi_q-1)^2}\nonumber\\
&\times& \int \frac{(1-v\sin\theta_k\cos\phi_k^0)^2 d(\cos\theta_k)}{|\sin\theta_k||\sin(\phi_k^0-\phi_q)|},
\label{32}
\end{eqnarray}
\begin{eqnarray}
\mathcal{L}(v)&=& \int^{1}_{-1}\frac{\sin^2\theta_q d(\cos\theta_q)}{|\sin\theta_q|}\int^{2\pi}_{0}\frac{\sin^2\phi_q d\phi_q}{(v^2\sin^2\theta_q\cos^2\phi_q-1)^2}\nonumber\\
&\times& \int \frac{(1-v\sin\theta_k\cos\phi_k^0)^2 d(\cos\theta_k)}{|\sin\theta_k||\sin(\phi_k^0-\phi_q)|}.
\label{33}
\end{eqnarray}

%{\color{blue}Discuss here the results for velocity dependent functions}
Eqs.~\ref{31},~\ref{32} and \ref{33} have been solved numerically and the relevant plots for the diffusion coefficients are presented in section~{\ref{results}}.
%%%%%%%%%%%%%%%%%%%%%%%%%%%%%%%%%%%%%%%%%%%%%%%%%%%%
\subsection{Quark contribution}  
\label{fermionic}
%%%%%%%%%%%%%%%%%%%%%%%%%%%%%%%%%%%%%%%%%%%%%%%%%%%%

The other contribution to the diffusion coefficients in the LLL approximation arises from the Coulomb scattering, $i.e.$, scattering of HQ with that of LLL light thermal quarks. Let us first consider the case in which HQ moves in the direction of the magnetic field with velocity $\textbf{v}$.  Here, we use kinetic theory approach similar to Ref.~\cite{Romatschke:2006bb} in which the momentum diffusion coefficient is related to energy loss per unit time, given as
\begin{equation}
\frac{dE}{dt}=\Re \int d^4Q j^{i}_{ext}(Q) E^{i}_{ind}(Q).
\label{enlosst}
\end{equation}	
Here $j^{i}_{ext}(Q)=2\pi q^{a}\textbf{v}\delta(\omega-\textbf{v}\cdot \textbf{q})$	is external quark current of chare $q^a$ moving with velocity $\textbf{v}$ and $E^{i}_{ind}$ is induced color electric field. In the non-relativistic limit or $q < T$,  the diffusion coefficient is obtained by using the relation $\kappa=- (2T/v^2) dE/dt$. Further, to obtain the induced chromo-electric field, one can solve Maxwell equation~\cite{Romatschke:thesis}
\begin{equation}
-i Q_{\mu} F^{\mu \nu}(Q)=j_{ind}^{\nu}(Q)+j^{\nu}_{ext}(Q),
\label{max}
\end{equation}
where $j_{ind}^{\mu}(Q)=A_{\nu}(Q)\Pi^{\mu \nu}(Q)$ is the induced current. Here $\Pi_{\mu \nu}(Q)$ is the gluon self energy in the presence of a constant magnetic field background. This self energy in the LLL approximation is proportional to $\alpha_s |eB|$; further, for leading order estimations of diffusion coefficient $F^{\mu \nu}=\partial_{\mu}A_{\nu}-\partial_{\nu}A_{\mu}$. Note here that in $F^{\mu \nu}$, the term containing the coupling $g$ is dropped for leading order calculations~\cite{Romatschke:thesis}. To obtain the induced electric field, defined as $E_{i}=i q_{i}A_{0}-i\omega A_{i}$, Eq.~(\ref{max}), can be simplified  in form
\begin{equation}
(D^{-1})^{00} A_{0}+(D^{-1})^{i0}A_{i}=-j^{0}_{ext},
\label{a1}
\end{equation}
\begin{equation}
(D^{-1})^{0k} A_{0}+(D^{-1})^{ik}A_{i}=-j^{k}_{ext},
\label{a2}
\end{equation}
where $j^{0}_{ext}=2\pi q^{a}\delta(\omega-\textbf{v}\cdot\textbf{q})$ and $(D^{-1})^{\mu \nu}(Q)=Q^2g^{\mu \nu}-Q^{\mu}Q^{\nu}+\Pi^{\mu \nu}(Q)$. Above two equation can be solved to obtain $A_{0}, A_{i}$ and hence $E_{i}$. Explicit forms of $A_{0}, A_{i}$ are given in appendix~\ref{maxeqn}. With $E_{i}=i q_{i}A_{0}-i\omega A_{i}$ and Eq.~(\ref{enlosst}), the quark contribution to the diffusion coefficient  $\kappa^{\parallel}_{LL; Qq}$ is given as
\begin{eqnarray}
\kappa^{\parallel}_{LL;Qq}=\frac{g^2(N^2-1)}{2N}\Im\int\frac{d^4Q}{(2\pi)^4}\frac{2Tq_z^2}{\omega}\bigg(v_{i}A_{i}-A_{0}\bigg).\nonumber\\
\label{kappall}
\end{eqnarray} 
Similarly, the component transverse to the magnetic field can be written as
\begin{eqnarray}
\kappa^{\parallel}_{TT;Qq}=\frac{g^2(N^2-1)}{2N}\Im\int\frac{d^4Q}{(2\pi)^4}\frac{2Tq_{\perp}^2}{\omega}\bigg(v_{i}A_{i}-A_{0}\bigg),\nonumber\\
\label{kappatt}
\end{eqnarray} 
where $q_{\perp}^2=q^2\sin^2\theta_q$. In the axial gauge ($i.e.$ $A_0=0$), the expressions for $\kappa^{\parallel}_{LL/TT}$ reduces to the one obtained in Ref.~\cite{Romatschke:2006bb}. Moreover, in this case one needs to write propagators in the same gauge.

Next, we consider the quark contribution to the diffusion coefficient for the case of HQ moving perpendicular to the magnetic field. Similar to Eqs.~(\ref{kappall}) and~(\ref{kappatt}), the other three diffusion coefficients are given as
\begin{eqnarray}
\kappa^{\perp}_{TL;Qq}=\frac{g^2(N^2-1)}{2N}\Im\int\frac{d^4Q}{(2\pi)^4}\frac{2Tq_z^2}{\omega}\bigg(v_{i}A_{i}-A_{0}\bigg),\nonumber\\
\label{kappatl}
\end{eqnarray} 

\begin{eqnarray}
\kappa^{\perp}_{LT;Qq}=\frac{g^2(N^2-1)}{2N}\Im\int\frac{d^4Q}{(2\pi)^4}\frac{2Tq_x^2}{\omega}\bigg(v_{i}A_{i}-A_{0}\bigg),\nonumber\\
\label{kappalt}
\end{eqnarray} 

\begin{eqnarray}
\kappa^{\perp}_{TT;Qq}=\frac{g^2(N^2-1)}{2N}\Im\int\frac{d^4Q}{(2\pi)^4}\frac{2Tq_y^2}{\omega}\bigg(v_{i}A_{i}-A_{0}\bigg).\nonumber\\
\label{kappattQq}
\end{eqnarray} 
An explicit form of the terms contributing to the diffusion coefficients is presented in appendix~\ref{maxeqn}. 

%%%%%%%%%%%%%%%%%%%%%%%%%%%%%%%%%%%%%%%%%%%%%%%%

\subsubsection*{Case I: $\textbf{v}\parallel\textbf{B}$} 
Since HQ velocity is along $\hat{z}$ direction, only $A_3$ and $A_0$ components contribute in the calculation of $\kappa_{LL/TT;Qq}^{\parallel}$. From Eqs.(\ref{a0}) and (\ref{ai}), it can be seen that there are various terms such as, e.g., for $i=1, k=2$, $\mathcal{D}^{12}(Q)\tilde{\Delta}(A)^{12}$. Here we shall simplify a few terms in $A_0$ and $A_3$, and other terms can also be simplified in a similar way. Let us consider the term $\mathcal{D}^{11}\tilde{\Delta}^{11}$ where the imaginary part can be written as
\begin{eqnarray}
\Im(\mathcal{D}^{11}\tilde{\Delta}^{11})=\frac{-(\omega^2-q^2+q_x^2)\sigma_{11}\delta(\omega-\textbf{v}\cdot\textbf{q})}{\gamma_{11}^2+\sigma_{11}^2},
\end{eqnarray}
where $\gamma_{11}=(Q^2+q_x^2)(q^2+q_z^2\Re \Pi_{\parallel})-\omega^2q_x^2$ and $\sigma_{11}=q_z^2(Q^2+q_x^2)\Im\Pi_{\parallel}$. After performing energy integration in Eq.(\ref{kappall}) with the delta function, the diffusion coefficient for this term can be written as
\begin{eqnarray}
\kappa_{LL}^{\parallel}|_{11}&=&-\frac{g^2 (N^2-1)}{2N}\int\frac{d\textbf{q}}{(2\pi)^3}\frac{2 Tq_z}{v}(v^2 q_z^2-q^2+q_x^2)\nonumber\\
&\times&\bigg[\frac{\sigma_{11}}{\gamma_{11}^2+\sigma_{11}^2}\bigg]_{\omega=vq_z}.
\label{kappa11}
\end{eqnarray}
Further, the imaginary part of the self energy, i.e., $\Im\Pi_{\parallel}$, given in Eq.(\ref{imagself}), can be written as $\Im\Pi_{\parallel}=\frac{\alpha_s|q_f eB|}{v q_z}\frac{2}{1-v^2}\exp\bigg({-\frac{q_{\perp}^2}{|q_f eB|}}\bigg)\delta(q_z)$
%{\color{red}separate both the contributions}. 
Note that in the limit that $q_{\perp}^2\ll |eB|$, the exponential factor is suppressed. Performing the $q_z$ integration using the delta function in the imaginary part of the self energy, the contribution in the diffusion coefficient vanishes, i.e., $\kappa_{LL}^{\parallel}|_{11}=0$. Similarly, it can be shown that the contribution from all other terms in $A_0$ and $A_3$ vanishes due to the delta function so that one obtains
\begin{equation}
\kappa_{LL;Qq}^{\parallel}=0.
\label{kappall1}
\end{equation}
This result is the same as for the case of static limit found in Ref.~\cite{Fukushima:2015wck}. It turns out that the vanishing of the longitudinal diffusion coefficient in the light quark massless limit is universal for both static and non-static limits.

\noindent
\textbf{Evaluation of $\kappa_{TT;Qq}^{\parallel}:$}  Now, let us estimate the contribution of $\mathcal{D}^{11}\tilde{\Delta}^{11}$ on the transverse component of the diffusion coefficient which after performing the energy integration using the  delta function becomes
\begin{eqnarray}
\kappa_{TT}^{\parallel}|_{11}&=&-\frac{g^2 (N^2-1)}{2N}\int\frac{d\textbf{q}}{(2\pi)^3}\frac{2 Tq_{\perp}^2}{vq_z}(v^2 q_z^2-q^2+q_x^2)\nonumber\\
&\times&\bigg[\frac{\sigma_{11}}{\gamma_{11}^2+\sigma_{11}^2}\bigg]_{\omega=vq_z}.
\end{eqnarray}
Similar to Eq.~(\ref{kappa11}), performing $q_z$ integration using the delta function, 
$\kappa_{TT}^{\parallel}|_{11}$ can be written as
\begin{equation}
\kappa_{TT}^{\parallel}|_{11}=-\frac{g^2 (N^2-1)}{2N}\frac{4 \alpha_s T|q_feB|}{v^2(1-v^2)}\int \frac{d^2q}{(2\pi)^2}\bigg[\frac{q_{\perp}^2 e^{-\frac{q_{\perp}^2}{|q_feB|}}}{ (q_{\perp}^2+\Re\Pi_{\parallel}^{0})^2}\bigg],
\label{kappatt1}
\end{equation}
where $\Re\Pi_{\parallel}^0=\frac{2\alpha_s|q_f eB|}{\pi (1-v^2)}e^{-\frac{q^2_{\perp}}{|q_feB|}}$. Note that unlike the contribution in longitudinal diffusion, the contribution of $\mathcal{D}^{11}\tilde{\Delta}^{11}$ in transverse diffusion does not vanish. Similarly, it can be shown that other components of $A_3$ and $A_0$ does not vanish so that one can obtain a finite value of transverse diffusion coefficient, i.e, $\kappa_{TT;Qq}^{\parallel} \propto T |q_f eB|\neq 0$. The contributions from the terms $\mathcal{D}^{12}\tilde{\Delta}^{12}, \mathcal{D}^{21}\tilde{\Delta}^{21},\mathcal{D}^{22}\tilde{\Delta}^{22}, \mathcal{D}^{31}\tilde{\Delta}^{31}$ and $\mathcal{D}^{32}\tilde{\Delta}^{32}$ are same as $\mathcal{D}^{11}\tilde{\Delta}^{11}$. The contributions from other terms are discussed below.   Let us note that in the limit $q_{\perp}^2\ll |q_f eB|$, the integral is logarithmically divergent. 
%{\color{red}write how to cure}. 
On the other hand, in the limit $q_{\perp}^2\sim |q_feB|$, the integral is convergent and reads as
\begin{eqnarray}
\int \frac{d^2q_{\perp}}{(2\pi)^2}\frac{q_{\perp}^2e^{-\frac{q_{\perp}^2}{2|q_feB|}}}{(q_{\perp}^2+\Re\Pi^{0}_{\parallel})^2}=-1-\gamma_E+\log\left(\frac{\pi(1-v^2)}{2\alpha_s}\right)\nonumber\\.
\label{logterm}
\end{eqnarray}
The converging nature of the integral naturally arises due to the finite size of LLL, i.e., $l\sim 1/\sqrt{|eB|}$. In fact, this is expected because the particles in LLL can not transfer momentum larger than $\sqrt{|eB|}$.

The contribution from the term $\mathcal{D}^{13}\tilde{\Delta}^{13}$ is 
\begin{eqnarray}
\kappa^{\parallel}_{TT}\bigg|_{13}=-\frac{4\alpha_s T|q_feB|(1-v^2)}{v^2}\int\frac{d^2q}{(2\pi)^2}\frac{q_{\perp}^2e^{-\frac{q_{\perp}^2}{2|q_feB|}}}{(q_{\perp}^2+\Re\Pi^{1})^2},\nonumber\\
\end{eqnarray}
where $\Re\Pi^{1}=2\alpha_s |q_feB|/\pi$. Contribution from the term $\mathcal{D}^{23}\tilde{\Delta}^{23}$ is same as $\mathcal{D}^{13}\tilde{\Delta}^{13}$. Finally, the contribution from the term $\mathcal{D}^{33}\tilde{\Delta}^{33}$ is given as
\begin{eqnarray}
\kappa^{\parallel}_{TT}\bigg|_{33}=-\frac{4\alpha_s T|q_feB|}{v}\int\frac{d^2q}{(2\pi)^2}\frac{q_{\perp}^2e^{-\frac{q_{\perp}^2}{2|q_feB|}}}{(q_{\perp}^2+\Re\Pi^{1})^2}.
\end{eqnarray}
In $A_3$, the terms $\mathcal{D}^{01}\tilde{\Delta}^{31}$ and $\mathcal{D}^{02}\tilde{\Delta}^{32}$ are same as $-v\mathcal{D}^{31}\tilde{\Delta}^{31}$. Furthermore, contribution from the term $\mathcal{D}^{03}\tilde{\Delta}^{33}$ is given as
\begin{eqnarray}
\kappa^{\parallel}_{TT}\bigg|_{03}=\frac{4\alpha_sT|q_feB|v}{v(1+v)}\int\frac{d^2q}{(2\pi)^2}\frac{q_{\perp}^2e^{-\frac{q_{\perp}^2}{2|q_feB|}}}{(q_{\perp}^2+\Re\Pi^{1})^2}.
\end{eqnarray}
Therefore, the final contribution to the diffusion coefficient  is
\begin{equation}
\kappa^{\parallel}_{TT}=\kappa^{\parallel}_{TT}\bigg|_{03}-2(1+v)\kappa^{\parallel}_{TT}\bigg|_{13}-6\kappa^{\parallel}_{TT}\bigg|_{11}-\kappa^{\parallel}_{TT}\bigg|_{33}.
\end{equation}

\subsubsection{Case II: $\textbf{v}\perp\textbf{B}$}
In this case, we assume that HQ velocity $\textbf{v}=v \hat{x}$, so that energy transfer $\omega=vq_x$. Again similar to $\textbf{v}\parallel\textbf{B}$ case, we shall evaluate only one component for each diffusion coefficient. Since the mathematical structure of other terms is quite the same as $\mathcal{D}^{11}\tilde{\Delta}^{11}$, it is easy to generalize the results. Let us start with $\kappa_{TL;Qq}^{\perp}$ which can be given as
\begin{equation}
\kappa_{TL}^{\perp}|_{11}=\frac{g^2 (N^2-1)}{2N}\int\frac{d^4Q}{(2\pi)^3}\frac{2 Tq_z^2}{\omega}\Im (\mathcal{D}^{11}\tilde{\Delta}^{11}).
\label{kappaTL11}
\end{equation}
where
\begin{eqnarray}
\Im(\mathcal{D}^{11}\tilde{\Delta}^{11})=-\frac{(Q^2+q_x^2-v\omega q_x)\rho}{\eta^2+\rho^2}.
\end{eqnarray}
Here, $\rho=q_z^2(Q^2+q_x^2)\Im\Pi_{\parallel}$ and $\eta=q^2(Q^2+q_x^2)-\omega^2q_x^2+q_z^2(Q^2+q_x^2)\Re\Pi_{\parallel}$. In order to simplify above equation, we write $\Im\Pi_{\parallel}=\frac{2\alpha_s|q_feB|}{2 \omega}(\delta(\omega-q_z)+\delta(\omega+q_z))$ and perform energy integration using delta function $\delta(\omega-\textbf{v}\cdot\textbf{q})$, and acquire
\begin{eqnarray}
\kappa_{TL}^{\perp}|_{11}&=&\frac{g^2 \alpha_sT|q_feB|(N^2-1)}{Nv^2}\int \frac{d^3q}{(2\pi)^3}\frac{q_z^2}{q_x^2}[\delta(vq_x-q_z)\nonumber\\
&+&\delta(vq_x+q_z)]\bigg(\frac{q_z^2((v^2+1)q_x^2-q^2)(q_z^2+q_y^2)}{\eta^2+\rho^2}\bigg).
\end{eqnarray}
The terms $\delta(vq_x\pm q_z)$ represent the LLL states along and opposite to the direction of the magnetic field. As may be noted, both the states give equal contribution to the diffusion coefficient. Therefore, we can use any delta function for $q_z$ integration and multiply overall quantity by a factor of two. After performing $q_z$ integration by using the delta function, we get
\begin{eqnarray}
\kappa_{TL}^{\perp}|_{11}&=&0.
\label{kappaTL}
\end{eqnarray}
Hence, $\mathcal{D}^{11}\tilde{\Delta}^{11}$ contribution to the $\kappa_{LT}^{\perp}$ vanishes. This can be realised from the real part of self energy $\Re\Pi_{\parallel}=\frac{\alpha_s |q_f eB|}{\pi q_{\parallel}^2}$.  The finite contribution comes only from $\mathcal{D}^{31}\tilde{\Delta}^{31}$ term in $A_0$ which is given as 
\begin{eqnarray}
\kappa_{TL}^{\perp}|_{31}&=&\frac{g^2 (N^2-1)}{2N}\int\frac{d^4Q}{(2\pi)^2}\frac{2 T q_z^2}{\omega}\nonumber\\
&\times&\bigg[\frac{(-v q_{\perp}^2+q_{\parallel}^2) \Im\Pi_{\parallel}\delta(\omega-\textbf{v}\cdot\textbf{q})}{(Q^2+q_{\parallel}^2\Re\Pi_{\parallel})^2+q_{\parallel}^4\Im\Pi_{\parallel}^2}\bigg].
\end{eqnarray} 
Performing the energy and $q_z$ integration using delta function one obtains
\begin{eqnarray}
\kappa_{TL}^{\perp}|_{31}&=&-\frac{2g^2v\alpha_sT|q_feB| (N^2-1)}{N}\int\frac{d^2q}{(2\pi)^2}\frac{q_{\perp}^2e^{-\frac{q_{\perp}^2}{2|q_feB|}}}{(q_{\perp}^2+\Re\Pi^1)^2}\nonumber\\ 
&\neq&0. 
\end{eqnarray}
For $q_{\perp}^2\sim |q_feB|$, the above integration can be performed similar to Eq.~(\ref{logterm}). Furthermore, two out of three terms arising from $A_{1}$ vanishes due to the delta function, and only $\mathcal{D}^{03}\tilde{\Delta}^{13}$ gives finite contribution. The imaginary part of this term reads as
\begin{eqnarray}
\Im(\mathcal{D}^{03}\tilde{\Delta}^{13})=\frac{\omega q_x q_{\perp}^2 \Im \Pi_{\parallel}\delta(\omega-\textbf{v}\cdot\textbf{q})}{\gamma_{03}^2+\sigma_{03}^2},
\end{eqnarray}
where
\begin{equation}
\gamma_{03}^2=q_x^2(Q^2+q_{\parallel}^2\Re\Pi_{\parallel})^2,\hspace{0.3cm} \sigma_{03}=q_x^2(q_{\parallel}^2\Im\Pi_{\parallel})^2.
\end{equation}
Therefore, $A_1$ contribution to diffusion coefficient can be written as
\begin{eqnarray}
\kappa^{\perp}_{TL}|^{A_1}=\frac{g^2(N^2-1)}{2N}\int\frac{d^4Q}{(2\pi)^4}\frac{2Tq_z^2}{\omega}\bigg[\frac{\omega q_x q_{\perp}^2\Im\Pi_{\parallel}\delta(\omega-\textbf{v}\cdot\textbf{q})}{\gamma_{03}^2+\sigma_{03}^2}\bigg].\nonumber\\ 
\end{eqnarray}
Simplifying the above equation similar to the previous one and performing $q_z$ integration using the delta function, one obtains
\begin{equation}
\kappa^{\perp}_{TL}|^{A_1}=\frac{2g^2v\alpha_sT|q_feB| (N^2-1)}{N} \int\frac{d^2 q}{(2\pi)^2}\frac{q_{\perp}^2e^{-\frac{q_{\perp}^2}{2|q_feB|}}}{(q_{\perp}^2+\Re\Pi^{1})^2}. 
\end{equation}
Adding both $A_0$ and $A_1$ contribution, the final form of the diffusion coefficient reads 
\begin{eqnarray}
\kappa^{\perp}_{TL;Qq}&=&v\kappa^{\perp}_{TL}|^{A_1}-\kappa_{TL}^{\perp}|_{31},\nonumber\\
&=&\frac{2g^2(v+v^2)\alpha_sT|q_feB| (N^2-1)}{N}\nonumber\\
&\times &\int\frac{d^2q}{(2\pi)^2}\frac{q_{\perp}^2e^{-\frac{q_{\perp}^2}{2|q_feB|}}}{(q_{\perp}^2+\Re\Pi^1)^2}.\nonumber\\
\end{eqnarray}

\noindent
\textbf{Evaluation of $\kappa^{\perp}_{LT;Qq}$:} The diffusion coefficient along the $\hat{x}$ direction is given in Eq.~(\ref{kappalt}). Let us first consider the terms arising from $A_0$. In this case the finite contribution comes from $\mathcal{D}^{31}\tilde{\Delta}^{31}$, the imaginary part of which reads as
\begin{eqnarray}
\Im(\mathcal{D}^{31}\tilde{\Delta}^{31})=\frac{(-vq_{\perp}^2+q_{\parallel}^2)\Im\Pi_{\parallel}\delta(\omega-\textbf{v}\cdot\textbf{q})}{\gamma_{31}^2+\sigma_{31}^2},
\end{eqnarray}
where
\begin{equation}
\gamma_{31}^2=(Q^2+q_{\parallel}^2\Re\Pi_{\parallel})^2, \hspace{0.3cm} \sigma_{31}^2=(q_{\parallel}^2\Im\Pi_{\parallel})^2. 
\end{equation}
Thus, the diffusion coefficient takes the form as follows,
\begin{eqnarray}
\kappa_{LT}^{\perp}|_{31}&=&\frac{g^2(N^2-1)}{2N}\int\frac{d^4Q}{(2\pi)^4}\frac{2 Tq_x^2}{\omega}\frac{(-vq_{\perp}^2+q_{\parallel}^2)\Im\Pi_{\parallel}}{\gamma_{31}^2+\sigma_{31}^2}.\nonumber\\ 
&\times&\delta(\omega-\textbf{v}\cdot\textbf{q}).
\end{eqnarray}
Further, using the energy delta function and momentum delta function of self energy, one can perform energy and $q_z$ integration to acquire
\begin{eqnarray}
\kappa_{LT}^{\perp}|_{31}=-\frac{2g^2T\alpha_s|q_feB|(N^2-1)}{2Nv}\int \frac{d^2q}{(2\pi)^2}\frac{q_{\perp}^2e^{-\frac{q_{\perp}^2}{2|q_feB|}}}{(q_{\perp}^2+\Re\Pi^{1}_{\parallel})^2}.\nonumber\\ 
\end{eqnarray}
Furthermore, the finite contribution in $A_{1}$ arises from the term $\mathcal{D}^{03}\tilde{\Delta}^{13}$ so that the diffusion coefficient reads as
\begin{eqnarray}
\kappa_{LT}^{\perp}|^{A_1}&=&\frac{g^2(N^2-1)}{2N}\int \frac{d^4Q}{(2\pi)^4}\frac{2Tq_x^2}{\omega}\nonumber\\
&\times&\bigg[\frac{\omega q_x q_{\perp}^2\Im\Pi_{\parallel}\delta(\omega-\textbf{v}\cdot\textbf{q})}{q_x^2[(Q^2+q_{\parallel}^2\Re\Pi_{\parallel})^2+(q_{\parallel}^2\Im\Pi_{\parallel})^2]}\bigg].
\end{eqnarray}
With further simplification, above equation can be written as
\begin{equation}
\kappa_{LT}^{\perp}|^{A_1}=\frac{2g^2T\alpha_s|q_feB|(N^2-1)}{2N}\int \frac{d^2q}{(2\pi)^2}\frac{q_{\perp}^2e^{-\frac{q_{\perp}^2}{2|q_feB|}}}{(q_{\perp}^2+\Re\Pi^{1})^2}.
\end{equation}
Thus, the total diffusion coefficient takes the form as
\begin{eqnarray}
\kappa_{LT;Qq}^{\perp}&=&v\kappa_{LT}^{\perp}|^{A_1}-\kappa_{LT}^{\perp}|_{31},\nonumber\\
&=&\frac{2g^2T\alpha_s|q_feB|(N^2-1)}{2N}\bigg(1+\frac{1}{v}\bigg)\nonumber\\
&\times&\int\frac{d^2q}{(2\pi)^2}\frac{q_{\perp}^2e^{-\frac{q_{\perp}^2}{2|q_feB|}}}{(q_{\perp}^2+\Re\Pi^{1})^2}.\nonumber\\
\end{eqnarray}

\noindent
\textbf{Evaluation of $\kappa_{TT;Qq}^{\perp}$:} Similar to $\kappa^{\perp}_{LT;Qq}$, the finite contribution arises from $\mathcal{D}^{31}\tilde{\Delta}^{31}$ and $\mathcal{D}^{03}\tilde{\Delta}^{13}$. Therefore, $\kappa^{\perp}_{TT;Qq}$ can be written as
\begin{eqnarray}
\kappa_{TT;Qq}^{\perp}&=&v\kappa_{TT}^{\perp}|^{A_1}-\kappa_{TT}^{\perp}|_{31}\nonumber\\
&=&\frac{2g^2T\alpha_s|q_feB|(N^2-1)}{2N}\bigg(1+\frac{1}{v}\bigg)\int\frac{d^2q}{(2\pi)^2}\nonumber\\
&\times& \bigg[\frac{q_{\perp}^2 \tan^2\phi e^{-\frac{q_{\perp}^2}{2|q_feB|}} }{(q_{\perp}^2+\Re\Pi^{1})^2}\bigg].
\end{eqnarray}
Futher, keeping in mind that $\omega>0$, one can separate the angular integration and simplify above integral as
\begin{eqnarray}
\kappa_{TT;Qq}^{\perp}&=&\frac{2\pi g^2T\alpha_s|q_feB|(N^2-1)}{2N}\bigg(1+\frac{1}{v}\bigg)\int \frac{dq_{\perp}^2}{(2\pi)^2}\nonumber\\
&\times&\bigg[\frac{q_{\perp}^2e^{-\frac{q_{\perp}^2}{|q_feB|}}  }{(q_{\perp}^2+\Re\Pi^{1})^2}\bigg].
\end{eqnarray}
%\bigg(\int_{0}^{\frac{\pi}{2}-\epsilon}d\phi\tan^2\phi \nonumber\\
%&+&\int_{\frac{3\pi}{2}+\epsilon}^{2\pi}d\phi\tan^2\phi\bigg)

\vspace{1 cm}

Now that the five components of diffusion have been estimated for both Coulomb and Compton scatterings of HQ, one can write down the total diffusion coefficients as
\begin{equation}
\kappa^{\parallel}_{LL}=\kappa^{\parallel}_{LL;Qg}+\kappa^{\parallel}_{LL;Qq},
\label{kappaparLLtot}
\end{equation} 
\begin{equation}
\kappa^{\parallel}_{TT}=\kappa^{\parallel}_{TT;Qg}+\kappa^{\parallel}_{TT;Qq},
\label{kappaparTTtot}
\end{equation}
\begin{equation}
\kappa^{\perp}_{LT}=\kappa^{\perp}_{LT;Qg}+\kappa^{\perp}_{LT;Qq},
\label{kappaperpLTtot}
\end{equation}
\begin{equation}
\kappa^{\perp}_{TL}=\kappa^{\perp}_{TL;Qg}+\kappa^{\perp}_{TL;Qq},
\label{kappaperpTLtot}
\end{equation}
\begin{equation}
\kappa^{\perp}_{TT}=\kappa^{\perp}_{TT;Qg}+\kappa^{\perp}_{TT;Qq}.
\label{kappaperpTTtot}
\end{equation}

\section{Effect of non-vanishing quark masses}
\label{non zero quark mass}
In the massive case, we consider the limits $m^2\ll q_{\parallel}^2$ as well as $\omega\ll T$. In the realistic situation of HIC, light quark mass is very small, i.e., $m\sim 5$ MeV, so that above limits in some sense are physical for QGP within the strong field limit. Further, in this limit, the leading mass term contribution to the imaginary part of self energy can be written as
\begin{equation}
\Im\Pi^{\parallel}_{R}(Q)=-\frac{g^2m^2 |q_feB|}{2q_{\parallel}^4}\frac{\beta \omega}{1+\cosh(\beta |q_{\parallel}|/2)}.
\label{img}
\end{equation} 
Furthermore, in the limit of small momentum transfer $q\sim g\sqrt{|eB|}\ll T$, one can further simplify the real part of self energy as
\begin{equation}
\Re\Pi_{\parallel}^R=\frac{2\alpha_s |eB|m^2}{T^2 q_{\parallel}^2}\int_{-\infty}^{\infty} \frac{dx}{(2\pi)}\frac{\tilde{f}(E)}{\sqrt{x^2+\hat{m}^2}}\bigg[\frac{1}{x^2+\hat{m}^2h(v)}\bigg],
\label{real}
\end{equation}
where $x=k_z/T, \hat{m}=m/T$ and $h(v)=v^2/(v^2-1)$. The above expression is obtained for a leading log approximation of the diffusion coefficient. For full leading contribution, one needs to take all other terms of self energy in account. It can clearly be seen in Eqs. (\ref{img}) and (\ref{real}) that the mass correction to the self energy and hence diffusion coefficients at leading order are suppressed by a factor $m^2/q_{\parallel}^2$. 
%%%%%%%%%%%%%%%%%%%%%%%%%%%%%%%%%%%%%%%%%%%%%%%%%%%%%% 
\section{Results and discussions}
\label{results}
%%%%%%%%%%%%%%%%%%%%%%%%%%%%%%%%%%%%%%%%%%%%%%%%%%%%%%%
First, we will report the results for the diffusion coefficients arising from Compton scattering of HQ.
\begin{figure}[h]
	\includegraphics[width=0.4\textwidth]{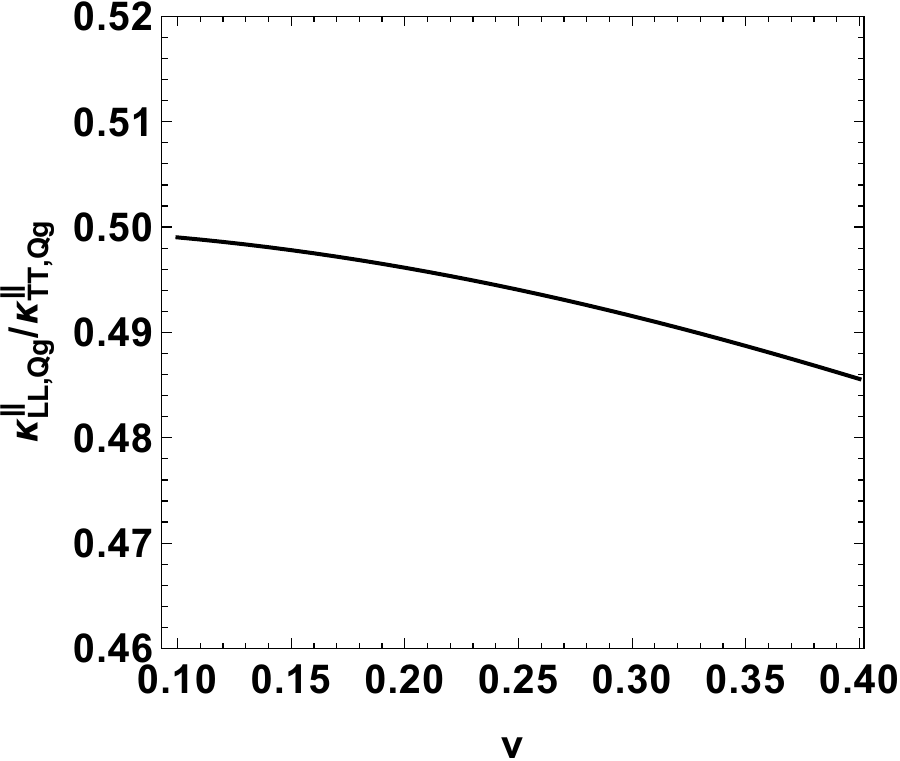}
	\caption{Ratio of the diffusion coefficients arising from Compton scattering process for a HQ moving in the direction of the magnetic field, $\it{i.e.}$ the z-axis, as a function of the velocity of HQ}
	\label{par}
\end{figure}
\begin{figure}[h]
	\includegraphics[width=0.4\textwidth]{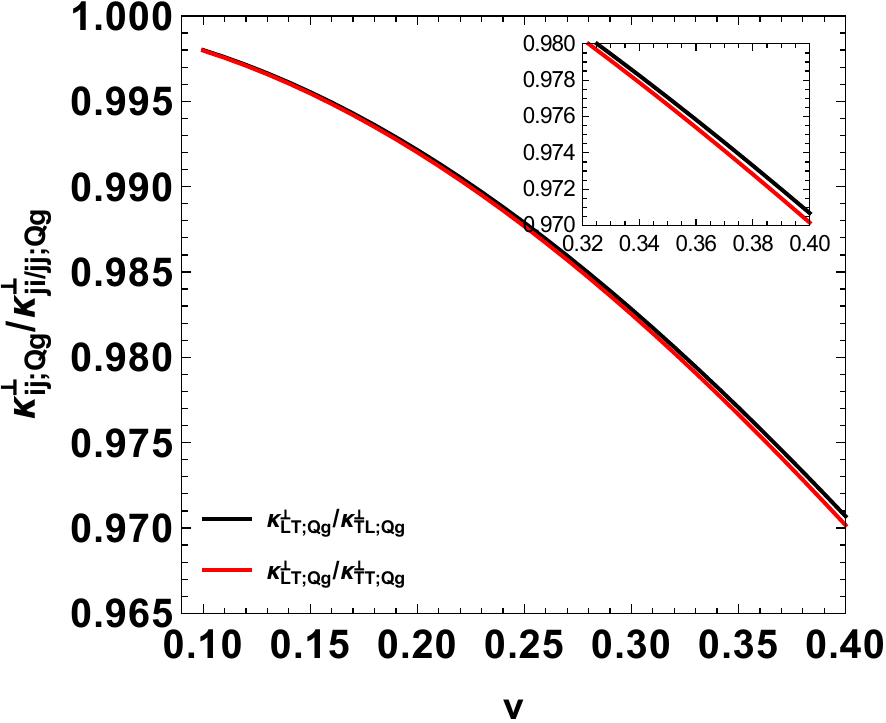}
	\caption{Ratios of the diffusion coefficients arising from the Compton scattering for a HQ moving perpendicular to the direction of the magnetic field(which is in z direction) as functions of the velocity of HQ. Here, $i=L$ and $j=T$ as appeared in the y-axis label}
	\label{perp}
\end{figure}
%{\color{red}Fig.(\ref{par}) and Fig.(\ref{perp}) are the velocity dependent angular integrated part of the five components of the diffusion coefficients.} 
For the case of HQ motion parallel to the magnetic field, we observe from Eq.~(\ref{23}) and Eq.~(\ref{24}) that the gluonic contributions are proportional to $T^3$ for Compton scattering.
We have plotted their ratio in Fig.\ref{par} which clearly indicates the dominance of $\kappa^{\parallel}_{TT;Qg}$, the diffusion component transverse to both $\textbf{v}$ and $\textbf{B}$, i.e. the z-axis, over $\kappa^{\parallel}_{LL;Qg}$, the diffusion longitudinal to the z-axis.
In case of the Coulomb scattering, $\kappa^{\parallel}_{LL;Qq}=0$ and $\kappa^{\parallel}_{TT;Qq}$ is proportional to $T|q_feB|$ as described in Eqs.~(\ref{kappall1}) and~(\ref{kappatt1}).
Thus, the ratio of these two diffusion coefficients satisfy, 
\begin{equation}
\frac{\kappa^{\parallel}_{LL}}{\kappa^{\parallel}_{TT}}\sim \frac{T^2}{eB}\ll 1.
\label{ratio}
\end{equation}
where $\kappa^{\parallel}_{LL}$ and $\kappa^{\parallel}_{TT}$, the total diffusion coefficients combining both Compton and Coulomb scattering, are given by Eqs.~\ref{kappaparLLtot} and \ref{kappaparTTtot}.
This observation is in line with that of Ref.~\cite{Fukushima:2015wck}, where the HQ diffusion coefficients are studied in a strong magnetic field at static limit. 
For the case of $\textbf{v}\perp\textbf{B}$, there are three diffusion coefficients, $\kappa^{\perp}_{LT}$, $\kappa^{\perp}_{TT}$ and $\kappa^{\perp}_{TL}$.
The ratios among these three coefficients are plotted in Fig.\ref{perp} which shows that both the ratios are comparable indicating the fact that all the three coefficients are similar to each other in magnitude.
The Coulomb scattering contribution to the diffusion coefficients is proportional to $T|q_feB|$. 
The calculations stipulate that the diffusion coefficients transverse to $\textbf{B}$, i.e. $\kappa^{\perp}_{LT;Qq}$ and $\kappa^{\perp}_{TT;Qq}$, are exactly same in magnitude and are dominant in comparison to the diffusion coefficient along the direction of $\textbf{B}$, i.e. $\kappa^{\perp}_{TL;Qq}$. As we found that Coulomb scattering contribution is larger compared to that of Compton scattering in the presence of the magnetic field, Coulomb scattering will dominate the nature of the ratios among the three diffusion coefficients.
Therefore, it can be inferred that according to Eqs.~\ref{kappaperpLTtot},~\ref{kappaperpTLtot} and \ref{kappaperpTTtot}, the total diffusion in the plane perpendicular to the magnetic field is dominant, i.e., $\kappa^{\perp}_{LT},\kappa^{\perp}_{TT}\gg \kappa^{\perp}_{TL}$. Therefore, the drag coefficient, related to the diffusion coefficient through the fluctuation-dissipation theorem, too, is larger in the transverse direction than in the longitudinal. The anisotropic HQ drag coefficients in the non-relativistic (NR) limit can be estimated by using the fluctuation-dissipation theorem as,
\begin{equation}
\eta^{\parallel}_{D}=\frac{\kappa^{\parallel}}{2 M T},
\label{43}
\end{equation} 
and 
\begin{equation}
\eta^{\perp}_{D}=\frac{\kappa^{\perp}}{2 M T}.
\label{44}
\end{equation} 
For the case of HQ moving parallel to the magnetic field, the drag coefficient perpendicular to the magnetic field is dominant, $i.e.$, $\eta^{\parallel}_{D;TT}>\eta^{\parallel}_{D;LL}$. This implies that HQ is dragged more in a plane transverse to the magnetic field (here, $xy$-plane) which may further generate anisotropic flow coefficients. Moreover, the relative magnitudes of the drag and the diffusion coefficients quantify the anisotropic nature of the transport coefficients. With an increase in the magnetic field in the LLL approximation, the drag/diffusion coefficients increase in magnitude, while the relative trend of the coefficients remains same. 

These results are only leading logarithmic and are strictly valid for a very small coupling, i.e., $g\ll 1$, small momentum transfer limit, i.e., $eB\gg |\textbf{q}|^2\gtrsim \alpha_s eB$ and in the regime $eB\gg T^2\gg \alpha_s eB$. Though the approximations used in this paper exclude many real effects of the HICs, it may be inferred that the various ratios among the components of the diffusion coefficients will remain unaltered, whatsoever. These results are the first step towards the understanding of anisotropic jet broadening in the presence of the magnetic field. 

All these results represent the case when light quark mass, $m$ is neglected all along. One may indicate some insights regarding the effect of the small mass of the light quarks. As we see from the expressions of the gluon self-energy from the Appendix~\ref{selfenergy}, there are terms proportional to $m^2$. In the current analysis, we dealt with the massless part of the gluon self energy to calculate the five components of the diffusion coefficients. Though the mass of light quark is much smaller than the other scales of the system, $i.e.$, $m\sim 5$ MeV, there will be a correction to the diffusion coefficients due to this finite quark mass. It is evident from the expressions that the relevant dimensionless ratio is $m^2/q_{\parallel}^2$ and in this case, $|\textbf{q}|^2\gtrsim\alpha_s eB$. Therefore, the correction will lie within the regime of scale, $m^2\ll \alpha_s eB$ so that the leading order(LO) mass-correction term will be proportional to $m^2/(\alpha_s eB)$. Let us take the case in which $\textbf{v}\parallel \textbf{B}$ as an example and observe the mass correction to the diffusion coefficients. Eq.~(\ref{amplparallel}) states that there are two terms proportional to $\Pi^{\parallel}_R$ and they can contribute in having terms of the order of $m^2$. The term proportional to
$q_{\parallel}^2\Pi^{\parallel}_R (P.K)$ contributes to the leading-log order and to the order $m^2$[Eq.~(\ref{25})]. Another term proportional to 
$\frac{q_{\parallel}^2\Pi^{\parallel}_R}{Q^2}(P.K)(K.Q)$ will not contribute to the leading-log but is of the order of $m^2$. All of the dimensionless ratios of $m^2$ with any other scales are much much less than 1 so that the massless leading logarithmic terms are always the dominant term within the realistic scales of the HICs.

%%%%%%%%%%%%%%%%%%%%%%%%%%%%%%%%%%%%%%%%%%%%%%%%%%%%
\section{summary and outlook}
\label{summary}
%%%%%%%%%%%%%%%%%%%%%%%%%%%%%%%%%%%%%%%%%%%%%%%%%%%%%%

The anisotropic diffusion and drag coefficients of HQ beyond the static limit have been computed in a constant magnetic field background. We have given the explicit expressions for the diffusion coefficients in the strong field approximation with the hierarchy $\alpha_seB\ll T^2\ll eB\ll M^2$ in the leading logarithmic approximation, $\it{i.e.}$ $log(T/m_{D,B})$. In the medium, the HQ makes multiple collisions with the thermal partons, $i.e.$, light quarks and gluons, and the process is akin to the Brownian motion. The magnetic field is assumed to be strong such that the condition $\sqrt{eB}\gg T$ is satisfied, and the dynamics of light quarks are restricted in the LLL. Further, it is also assumed that $M\gg\sqrt{eB}$, so that HQ is not directly affected by the magnetic field. To study the diffusion of HQ, the momentum transfer in the collision of HQ and the thermal partons is assumed to be small.

It is observed that one can define five independent momentum diffusion coefficients of HQ in the medium depending on the relative orientation of the volocity of HQ and magnetic field. In the case of HQ moving along the direction of the magnetic field, $\bf{v}\parallel \bf{B}$, the coefficients $\kappa^{\parallel}_{LL}$ and $\kappa^{\parallel}_{TT}$ quantify the momentum diffusion along ${\hat{z}}$ direction and transverse direction, i.e., $xy$ plane. Out of these two, diffusion along the direction of the magnetic field is smaller than the diffusion transverse to the direction of the magnetic field, $i.e.$,
\begin{equation}
\frac{\kappa^{\parallel}_{LL}}{\kappa^{\parallel}_{TT}}\ll 1.
\end{equation}
In the massless light quark limit, the contribution of the Coulomb scattering of HQ to the total diffusion coefficient, $\it{i.e.}$ $\kappa^{\parallel}_{LL}$ vanishes. This is because there is no momentum transfer along $\hat{z}$ direction. However, the thermal gluon contribution is proportional to $T^3$. Furthermore, light quark scattering contribution to $\kappa^{\parallel}_{TT}$ is proportional to $T|eB|$. Similarly, there are three diffusion coefficients for the case of HQ moving transverse to the direction of the magnetic field $i.e.$, $\bf{v}\perp \bf{B}$ denoted as $\kappa^{\perp}_{TL}, \kappa^{\perp}_{LT}, \kappa^{\perp}_{TT}$. It is observed that in this case, the drag coefficient in a plane transverse to the magnetic field is dominant.

Furthermore, the anisotropic transport coefficients can be used as input parameters for the estimation of HQ flow coefficients in the magnetized medium. It may be noted that HQ directed flow, $v_1$, is identified as a novel observable to probe the initial electromagnetic field produced in high energy collisions. The recent LHC measurement~\cite{Acharya:2019ijj},  along with the RHIC findings~\cite{Adam:2019wnk}, on the D-meson flow coefficient, $v_1$,  give the indications of the strong electromagnetic field  produced in high energy heavy-ion collisions. However, to compute the HQ directed flow, one needs to take into account the effect of electromagnetic field on HQ transport coefficients as well, which has been ignored in the previous calculations~\cite{Das:2016cwd,Chatterjee:2018lsx}. Heavy meson nuclear suppression factor and elliptic flow are the other experimentally measured observables that can be affected by the anisotropic HQ transport coefficient due to the presence of the electromagnetic field. The present investigation is limited to the strong and constant magnetic field case so that $eB\gg T^2$ where light quarks occupy only the LLL. For the case of the magnetic field of the order of the temperature $i.e.$, $eB\sim T^2$, higher Landau levels may give significant contributions to the transport coefficients. We intend to explore these aspects in the near future.

%%%%%%%%%%%%%%%%%%%%%%%%%%%%%%%%%%%%%%%%%%%%%%%%%%%%%%%%%%%%%%%%%
\section*{acknowledgments}

S.M. acknowledges IIT Gandhinagar for the academic visit and hospitality during the course of this work.
S.M is grateful to Dr. Jane Alam for his insightful discussions.
We are indebted to the people of India for their generous support for research in basic sciences.

%%%%%%%%%%%%%%%%%%%%%%%%%%%%%%%%%%%%%%%%%%%%%%%%%%%%%%%%%%%%%%%
\appendix

\section{Gluon self energy}
\label{selfenergy}
In the LLL approximation, the contribution of the gluon loop to the gluon self energy is proportional to $T^2$, and that of the quark loop is proportional to $eB$~\cite{Singh:2020fsj}. Hence, in the LLL approximation, $\it{i.e.}$ $eB\gg T^2$, the gluon loop contribution ($\sim T^2$) can be neglected with respect to the quark loop contribution ($\sim eB$). Assuming that the magnetic field $\textbf{B}$ is along the positive $\textbf{z}$-axis, the gauge-invariant structure of the gluon self energy is given as~\cite{Hattori:2012ny}
\begin{equation}
\Pi_{R}^{\mu \nu}(P)=\Pi_{R}^{0}(P)P_{0}^{\mu \nu}+\Pi_{R}^{\parallel}(P)P_{\parallel}^{\mu \nu}+\Pi_{R}^{\perp}(P)P_{\perp}^{\mu \nu},
\label{self}
\end{equation} 
where $P_{0,\parallel,\perp}^{\mu \nu}$ are the projection operators defined as,
\begin{equation}
P_{0}^{\mu \nu}=-P^2 g^{\mu \nu}+{P^{\mu}P^{\nu}}, P_{\parallel,\perp}^{\mu \nu}=-p_{\parallel,\perp}^2 g_{\parallel,\perp}^{\mu \nu}+{p_{\parallel,\perp}^{\mu}p_{\parallel,\perp}^{\nu}}.
\end{equation}
As in Eq.~(\ref{self}), $\Pi^{0}_{R}$ is vacuum term which is independent of both $T$ and $\textbf{B}$, we shall drop this term here. In the LLL, the light quark motion is restricted along the direction of the magnetic field only so that $\Pi^{\perp}_{R}(P)=0$. Further, the term $\Pi_{R}^{\parallel}=\Pi_{R}^{\parallel}(\textbf{B},T=0)+\Pi_{R}^{\parallel}(\textbf{B},T)$ where  $\Pi_{R}^{\parallel}(\textbf{B},T=0)$ depends on the magnetic field only and $\Pi_{R}^{\parallel}(\textbf{B},T)$ depends on both magnetic field and temperature. $\Pi^{R}_{\parallel}(\textbf{B},T=0)$ is given as~\cite{Hattori:2012ny}
\begin{equation}
\Pi_{R}^{\parallel}(\textbf{B},T=0)=\frac{\alpha_s |q_f eB|}{\pi [(\omega+i\epsilon)^2-p_z^2]}e^{-\frac{p_{\perp}^2}{|2q_f eB|}}\bigg(\mathcal{I}(\textbf{B})-2\bigg),
\label{self3}
\end{equation}
where, 
{\small
\begin{align}
\mathcal{I}(\textbf{B}) = \left\{ \begin{array}{cc} 
\frac{4m^2}{\sqrt{p_{\parallel}^2(p_{\parallel}^2-4m^2)}}\ln \bigg|\frac{p_{\parallel}^2-\sqrt{p_{\parallel}^2(p_{\parallel}^2-4m^2)}}{p_{\parallel}^2+\sqrt{p_{\parallel}^2(p_{\parallel}^2-4m^2)}}\bigg| & \hspace{0. mm} p_{\parallel}^2 < 0 \\
\frac{8m^2}{\sqrt{p_{\parallel}^2(4m^2-p_{\parallel}^2)}} \arctan\bigg(\frac{p_{\parallel}^2}{\sqrt{p_{\parallel}^2(4m^2-p_{\parallel}^2)}}\bigg) & \hspace{0. mm} 0<p_{\parallel}^2<4 m^2 \\
\frac{4m^2}{\sqrt{p_{\parallel}^2(p_{\parallel}^2-4m^2)}}\bigg[\ln \bigg|\frac{p_{\parallel}^2-\sqrt{p_{\parallel}^2(p_{\parallel}^2-4m^2)}}{p_{\parallel}^2+\sqrt{p_{\parallel}^2(p_{\parallel}^2-4m^2)}}\bigg|+i\pi\bigg] & \hspace{0. mm} p_{\parallel}^2 > 4m^2 \\
\end{array} \right.
\label{self2}
\end{align}}
Here, $m$ is mass of light quark. The first term  $\mathcal{I}(B)$ in Eq.~(\ref{self3}) corresponds to the mass correction and the second constant term represents the massless case.  Note that at $p_{\parallel}^2=4m^2$, $\mathcal{I}(B)$ has a singular behaviour. This singularity corresponds to the threshold for decay of gauge boson in a fermion-antifermion pair. Moreover, this can also be observed from the imaginary part in the self energy for $p_{\parallel}^2> 4m^2$. Furthermore, in the LLL approximation, the threshold does not depend on the strength of the magnetic field.  Another term that depends on $T$ and $\textbf{B}$, i.e., $\Pi^{\parallel}_{R}(\textbf{B},T)$  is given as
\begin{eqnarray}
\Pi^{\parallel}_{R}(\textbf{B},T)=\frac{\pi \Omega m^2}{p_{\parallel}^2} \bigg[\mathcal{J}_{0}(P)+ \frac{2 p_{z}}{p_{\parallel}^2}\mathcal{J}_{1}(P)\bigg],
\label{pir1}
\end{eqnarray} 
where 
\begin{equation}
\mathcal{J}_{a}=\int_{-\infty}^{\infty}\frac{dk_z}{2\pi E}\frac{\tilde{f}(E)k_z^{a}}{(k_z-p_z/2)^2-p_0^2/4+p_0^2 m^2/p_{\parallel}^2-i p_0 \epsilon},
\end{equation}
with
\begin{equation}
\Omega=\frac{\alpha|q_feB|}{2}\exp\bigg({-\frac{p_{\perp}^2}{|2q_feB|}}\bigg).
\end{equation}
Here, $q_f=1/3,2/3$ for light quark flavor $N_f=2$. For the massless case, the imaginary part of the self energy can be obtained from Eq.~(\ref{self3}) as
\begin{equation}
\Im \Pi^{\parallel}_{R}=2\alpha_s |q_feB|e^{-\frac{p_{\perp}^2}{|q_f eB|}}\delta(p_{\parallel}^2)
\label{imagself}
\end{equation}
Temperature dependent term in self energy as given by Eq.~(\ref{pir1}) vanishes in the massless limit. Hence, the contribution arises from the magnetic field only. On the other hand, for finite mass, as mentioned earlier, the imaginary part arises above fermion-antifermion threshold limit as shown in Eq.~(\ref{self2}). For the thermal part, the imaginary part of the self energy can be written as
% {\color{red}cite Baier}
\begin{eqnarray}
\Im \Pi_{R}^{\parallel}&=&-\frac{2 m^2 g^2|q_f eB|}{2 p_{\parallel}^2}e^{-\frac{p_{\perp}^2}{2|q_feB|}}\sinh\bigg(\frac{\beta \omega}{2}\bigg)\bigg[\int_{-\infty}^{\infty}\nonumber\\
&\times&\frac{dk_z}{\omega_{k_z}}\epsilon^0(\omega+\omega_{k_z})\delta(p_{\parallel}^2+2\omega \omega_{k_z}-2 k_z  p_z)e^{\beta \omega_{k_z}/2}\nonumber\\
&\times&e^{\beta |\omega+\omega_{k_z}|/2}\tilde{f}(\omega_{k_z})\tilde{f}(\omega+\omega_{k_z})-\int_{-\infty}^{\infty}\frac{dk_z}{\omega_{k_z}} \nonumber\\
&\times&\epsilon^{0}(\omega-\omega_{k_z})\delta(p_{\parallel}^2-2\omega \omega_{k_z}-2k_z p_z)e^{\beta \omega_{k_z}/2}\nonumber\\
&\times&e^{\beta |\omega-\omega_{k_z}|/2}\tilde{f}(\omega_{k_z})\tilde{f}(\omega-\omega_{k_z})
\bigg]
\end{eqnarray}
where $\epsilon^0$ is sign function, $\omega_{k_z}=\sqrt{k_z^2+m^2}$ and $\tilde{f}(E)$ is Fermi-Dirac distribution function. The integration can be solved using the delta function with simplification
\begin{eqnarray}
\delta(p_{\parallel}^2+2\omega \omega_{k_z}-2k_{z}p_{z})&=&\frac{\delta(k_z-k_z^0)}{|\frac{2\omega k_z}{\omega_{k_z}}-2 p_z|_{k_z=k_z^0}}\nonumber\\
&+&\frac{\delta(k_z-k_z^1)}{|\frac{2\omega k_z}{\omega_{k_z}}-2 p_z|_{k_z=k_z^1}}
\end{eqnarray} 
where
\begin{equation}
k_z^{0/1}=-\frac{p_z}{2}\pm\frac{\omega}{2}\sqrt{1-\frac{4m^2}{p_{\parallel}^2}}.
\label{pole}
\end{equation}
From Eq.~(\ref{pole}), it can clearly be seen that the contribution to imaginary part of self energy comes from two regions. One is fermion-antifermion threshold, i.e., $4m^2/p_{\parallel}<1$ and another is Landau damping, i.e., $p_{\parallel}^2 < 0$. With further simplifications, the imaginary part of the self energy can be written as
\begin{eqnarray}
\Im \Pi^{\parallel}_R&=&-\frac{m^2g^2|q_f eB|}{p_{\parallel}^4}e^{-\frac{p_{\perp}^2}{2|q_feB|}}\bigg(1-\frac{4m^2}{p_{\parallel}^2}\bigg)^{-1/2}\nonumber\\
&\times&\frac{\sinh(\frac{\beta \omega}{2})}{\cosh(\frac{\beta \omega}{2})+\cosh\bigg(\frac{\beta |p_{\parallel}|}{2}\sqrt{1-\frac{4m^2}{p_{\parallel}^2}}\bigg)}.
\end{eqnarray}
For the case of static limit, the imaginary part of the self energy is also evaluated in Ref.~(\cite{Singh:PRD972018}).

\section{$A_{0}$, $A_{i}$ terms for $\textbf{v}\parallel\textbf{B}$}
\label{maxeqn}
In terms of current $j^{\mu}$ and the propagators, $A_{0}$ can be obtained from Eqs.(\ref{a1}) and (\ref{a2}) as
\begin{equation}
A_{0}(Q)=\sum_{i,k=1}^{3} \mathcal{D}^{ik}(Q)\tilde{\Delta}^{ik}(Q)
\label{a0}
\end{equation}
where
\begin{equation}
\mathcal{D}^{ik}(Q)=(D^{-1})^{i0}(Q)j^{k}-(D^{-1})^{ik}(Q)j^{0}
\end{equation}
\begin{equation}
\tilde{\Delta}^{ik}(Q)=\frac{1}{(D^{-1})^{00}(Q)(D^{-1})^{ik}(Q)-(D^{-1})^{i0}(Q)(D^{-1})^{0k}(Q)}.
\end{equation}
Similary, $A_{i}$ is
\begin{equation}
A_{i}=\sum_{k=1}^{3}\mathcal{D}^{0k}(Q)\tilde{\Delta}^{ik}(Q)
\label{ai}
\end{equation}
where
\begin{equation}
\mathcal{D}^{0k}(Q)=(D^{-1})^{0k}(Q)j^{0}-(D^{-1})^{00}(Q)j^{k}.
\end{equation}
Below we give explicit forms of the terms summed in $A_0$ and drop the momentum argument $Q$ in $\mathcal{D}$ and $j$. Let us note that for $\textbf{v}\parallel\textbf{B}$ the currents $j^{1}=j^{2}=0$ and $j^{3}=2\pi q^{a}v\delta(\omega-\textbf{v}\cdot\textbf{q})$. The explicit terms are
\begin{eqnarray}
\mathcal{D}^{11}\tilde{\Delta}^{11}&=&\frac{(D^{-1})^{10}j^{1}-(D^{-1})^{11}j^{0}}{(D^{-1})^{00}(D^{-1})^{11}-(D^{-1})^{10}(D^{-1})^{01}}\nonumber\\
&=& \frac{(Q^2+q_x^2)2\pi\delta(\omega-\textbf{v}\cdot\textbf{q})}{(q^2+q_z^2\Pi_{\parallel})(Q^2+q_x^2)-\omega^2 q_x^2}
\end{eqnarray}
where $\Pi_{\parallel}$ is dimensionless quantity related to gluon self energy as $\Pi_{\parallel}=\Pi^{\parallel}_{R}$. Similarly, other terms are
\begin{eqnarray}
\mathcal{D}^{12}\tilde{\Delta}^{12}&=&\frac{(D^{-1})^{10}j^{2}-(D^{-1})^{12}j^{0}}{(D^{-1})^{00}(D^{-1})^{12}-(D^{-1})^{10}(D^{-1})^{02}}\nonumber\\
&=&\frac{2\pi\delta(\omega-\textbf{v}\cdot\textbf{q})}{q^2-\omega^2+q_z^2\Pi_{\parallel}}
\end{eqnarray}
\begin{eqnarray}
\mathcal{D}^{13}\tilde{\Delta}^{13}&=&\frac{(D^{-1})^{10}j^{3}-(D^{-1})^{13}j^{0}}{(D^{-1})^{00}(D^{-1})^{13}-(D^{-1})^{10}(D^{-1})^{03}}\nonumber\\
&=&\frac{q_z(1-v^2)2\pi\delta(\omega-\textbf{v}\cdot\textbf{q})}{q_z(q^2+q_z^2\Pi_{\parallel})-\omega^2q_z-\omega^2q_z\Pi_{\parallel}}
\end{eqnarray}
\begin{eqnarray}
\mathcal{D}^{21}\tilde{\Delta}^{21}&=&\frac{(D^{-1})^{20}j^{1}-(D^{-1})^{21}j^{0}}{(D^{-1})^{00}(D^{-1})^{21}-(D^{-1})^{20}(D^{-1})^{01}}\nonumber\\
&=&\frac{2\pi\delta(\omega-\textbf{v}\cdot\textbf{q})}{q^2+q_z^2\Pi_{\parallel}-\omega^2}
\end{eqnarray}
\begin{eqnarray}
\mathcal{D}^{22}\tilde{\Delta}^{22}&=&\frac{(D^{-1})^{20}j^{2}-(D^{-1})^{22}j^{0}}{(D^{-1})^{00}(D^{-1})^{22}-(D^{-1})^{20}(D^{-1})^{02}}\nonumber\\
&=&\frac{(Q^2+q_y^2)2\pi\delta(\omega-\textbf{v}\cdot\textbf{q})}{(Q^2+q_y^2)(q^2+q_z^2\Pi_{\parallel})-\omega^2q_y^2}
\end{eqnarray}
\begin{eqnarray}
\mathcal{D}^{23}\tilde{\Delta}^{23}&=&\frac{(D^{-1})^{20}j^{3}-(D^{-1})^{23}j^{0}}{(D^{-1})^{00}(D^{-1})^{23}-(D^{-1})^{20}(D^{-1})^{03}}\nonumber\\
&=&\frac{(q_yq_z-v\omega q_y)2\pi\delta(\omega-\textbf{v}\cdot\textbf{q})}{q_yq_z(q^2+q_z^2\Pi_{\parallel})-\omega^2 q_yq_z(1+\Pi_{\parallel})}
\end{eqnarray}
\begin{eqnarray}
\mathcal{D}^{31}\tilde{\Delta}^{31}&=&\frac{(D^{-1})^{30}j^{1}-(D^{-1})^{31}j^{0}}{(D^{-1})^{00}(D^{-1})^{31}-(D^{-1})^{30}(D^{-1})^{01}}\nonumber\\
&=&\frac{2\pi\delta(\omega-\textbf{v}\cdot\textbf{q})}{(q^2+q_z^2\Pi_{\parallel})-\omega^2 (1+\Pi_{\parallel})}
\end{eqnarray}
\begin{eqnarray}
\mathcal{D}^{32}\tilde{\Delta}^{32}&=&\frac{(D^{-1})^{30}j^{2}-(D^{-1})^{32}j^{0}}{(D^{-1})^{00}(D^{-1})^{32}-(D^{-1})^{30}(D^{-1})^{02}}\nonumber\\
&=&\frac{2\pi\delta(\omega-\textbf{v}\cdot\textbf{q})}{(q^2+q_z^2\Pi_{\parallel})-\omega^2 (1+\Pi_{\parallel})}
\end{eqnarray}

\begin{eqnarray}
\mathcal{D}^{33}\tilde{\Delta}^{33}=\frac{(D^{-1})^{30}j^{3}-(D^{-1})^{33}j^{0}}{(D^{-1})^{00}(D^{-1})^{33}-(D^{-1})^{30}(D^{-1})^{03}}\nonumber\\
=\frac{(Q^2+q_z^2-v\omega q_z+(\omega^2-v\omega q_z)\Pi_{\parallel})2\pi\delta(\omega-\textbf{v}\cdot\textbf{q})}{(Q^2+q_z^2+\omega^2\Pi_{\parallel})(q^2+q_z^2\Pi_{\parallel})-(\omega q_z+\omega q_z\Pi_{\parallel})^2}.\nonumber\\
\end{eqnarray}
Since velocity in along $\hat{z}$ direction, only $A_3$ component contribute to the diffusion coefficients and is given as
\begin{equation}
A_3=\mathcal{D}^{01}\tilde{\Delta}^{31}+\mathcal{D}^{02}\tilde{\Delta}^{32}+\mathcal{D}^{03}\tilde{\Delta}^{33}.
\label{a3}
\end{equation}
The explicit forms of the terms in Eq.~(\ref{a3}) are 
\begin{eqnarray}
\mathcal{D}^{01}\tilde{\Delta}^{31}&=&\frac{(D^{-1})^{01}j^{0}-(D^{-1})^{00}j^{1}}{(D^{-1})^{00}(D^{-1})^{31}-(D^{-1})^{30}(D^{-1})^{01}}\nonumber\\
&=&\frac{-\omega 2\pi\delta(\omega-\textbf{v}\cdot\textbf{q})}{q_z(q^2+q_z^2\Pi_{\parallel})-\omega^2q_z(1+\Pi_{\parallel})}
\end{eqnarray}
\begin{eqnarray}
\mathcal{D}^{02}\tilde{\Delta}^{32}&=&\frac{(D^{-1})^{02}j^{0}-(D^{-1})^{00}j^{2}}{(D^{-1})^{00}(D^{-1})^{32}-(D^{-1})^{30}(D^{-1})^{02}}\nonumber\\
&=&\frac{-\omega 2\pi\delta(\omega-\textbf{v}\cdot\textbf{q})}{q_z(q^2+q_z^2\Pi_{\parallel})-\omega^2q_z(1+\Pi_{\parallel})}
\end{eqnarray}
\begin{eqnarray}
\mathcal{D}^{03}\tilde{\Delta}^{33}=\frac{(D^{-1})^{03}j^{0}-(D^{-1})^{00}j^{3}}{(D^{-1})^{00}(D^{-1})^{33}-(D^{-1})^{30}(D^{-1})^{03}}\nonumber\\
=\frac{((v^2q^2-\omega q_z)+(v^2q_z^2-\omega q_z)\Pi_{\parallel})2\pi\delta(\omega-\textbf{v}\cdot\textbf{q})}{(Q^2+q_z^2)q^2-\omega^2q_z^2+(\omega^2q^2+q_z^2(Q^2+q_z^2)-2\omega^2q_z^2)\Pi_{\parallel}}\nonumber\\
\end{eqnarray}

\section{$A_{0}$, $A_{i}$ terms for $\textbf{v}\perp\textbf{B}$}
In this case, we assume that HQ is moving along $\hat{x}$ direction so that for small momentum transfer $\omega=vq_x$. Similar to the case of $\textbf{v}\parallel\textbf{B}$, the explicit terms in $A_0$ are
\begin{eqnarray}
\mathcal{D}^{11}\tilde{\Delta}^{11}&=&\frac{(D^{-1})^{10}j^{1}-(D^{-1})^{11}j^{0}}{(D^{-1})^{00}(D^{-1})^{11}-(D^{-1})^{10}(D^{-1})^{01}}\nonumber\\
&=& \frac{(Q^2+q_x^2-v\omega q_x )2\pi\delta(\omega-\textbf{v}\cdot\textbf{q})}{(Q^2+q_x^2)(q^2+q_z^2\Pi_{\parallel})-\omega^2 q_x^2}
\end{eqnarray}

\begin{eqnarray}
\mathcal{D}^{12}\tilde{\Delta}^{12}&=&\frac{(D^{-1})^{10}j^{2}-(D^{-1})^{12}j^{0}}{(D^{-1})^{00}(D^{-1})^{12}-(D^{-1})^{10}(D^{-1})^{02}}\nonumber\\
&=&\frac{2\pi\delta(\omega-\textbf{v}\cdot\textbf{q})}{q^2+q_z^2\Pi_{\parallel}-\omega^2}
\end{eqnarray}
\begin{eqnarray}
\mathcal{D}^{13}\tilde{\Delta}^{13}&=&\frac{(D^{-1})^{10}j^{3}-(D^{-1})^{13}j^{0}}{(D^{-1})^{00}(D^{-1})^{13}-(D^{-1})^{10}(D^{-1})^{03}}\nonumber\\
&=&\frac{2\pi\delta(\omega-\textbf{v}\cdot\textbf{q})}{(q^2+q_z^2\Pi_{\parallel})-\omega^2(1+\Pi_{\parallel})}
\end{eqnarray}
\begin{eqnarray}
\mathcal{D}^{21}\tilde{\Delta}^{21}&=&\frac{(D^{-1})^{20}j^{1}-(D^{-1})^{21}j^{0}}{(D^{-1})^{00}(D^{-1})^{21}-(D^{-1})^{20}(D^{-1})^{01}}\nonumber\\
&=&\frac{(1-v^2)2\pi\delta(\omega-\textbf{v}\cdot\textbf{q})}{q^2+q_z^2\Pi_{\parallel}-\omega^2}
\end{eqnarray}
\begin{eqnarray}
\mathcal{D}^{22}\tilde{\Delta}^{22}&=&\frac{(D^{-1})^{20}j^{2}-(D^{-1})^{22}j^{0}}{(D^{-1})^{00}(D^{-1})^{22}-(D^{-1})^{20}(D^{-1})^{02}}\nonumber\\
&=&\frac{(Q^2+q_y^2)2\pi\delta(\omega-\textbf{v}\cdot\textbf{q})}{(Q^2+q_y^2)(q^2+q_z^2\Pi_{\parallel})-\omega^2q_y^2}
\end{eqnarray}
\begin{eqnarray}
\mathcal{D}^{23}\tilde{\Delta}^{23}&=&\frac{(D^{-1})^{20}j^{3}-(D^{-1})^{23}j^{0}}{(D^{-1})^{00}(D^{-1})^{23}-(D^{-1})^{20}(D^{-1})^{03}}\nonumber\\
&=&\frac{2\pi\delta(\omega-\textbf{v}\cdot\textbf{q})}{q^2+q_z^2\Pi_{\parallel}-\omega^2(1+\Pi_{\parallel})}
\end{eqnarray}
\begin{eqnarray}
\mathcal{D}^{31}\tilde{\Delta}^{31}&=&\frac{(D^{-1})^{30}j^{1}-(D^{-1})^{31}j^{0}}{(D^{-1})^{00}(D^{-1})^{31}-(D^{-1})^{30}(D^{-1})^{01}}\nonumber\\
&=&\frac{(1-v-v\Pi_{\parallel})2\pi\delta(\omega-\textbf{v}\cdot\textbf{q})}{(q^2+q_z^2\Pi_{\parallel})-\omega^2 (1+\Pi_{\parallel})}
\end{eqnarray}
\begin{eqnarray}
\mathcal{D}^{32}\tilde{\Delta}^{32}&=&\frac{(D^{-1})^{30}j^{2}-(D^{-1})^{32}j^{0}}{(D^{-1})^{00}(D^{-1})^{32}-(D^{-1})^{30}(D^{-1})^{02}}\nonumber\\
&=&\frac{2\pi\delta(\omega-\textbf{v}\cdot\textbf{q})}{(q^2+q_z^2\Pi_{\parallel})-\omega^2 (1+\Pi_{\parallel})}
\end{eqnarray}
\begin{eqnarray}
\mathcal{D}^{33}\tilde{\Delta}^{33}=\frac{(D^{-1})^{30}j^{3}-(D^{-1})^{33}j^{0}}{(D^{-1})^{00}(D^{-1})^{33}-(D^{-1})^{30}(D^{-1})^{03}}\nonumber\\
=\frac{(Q^2+q_z^2+\omega^2)2\pi\delta(\omega-\textbf{v}\cdot\textbf{q})}{(Q^2+q_z^2+\omega^2\Pi_{\parallel})(q^2+q_z^2\Pi_{\parallel})-(\omega q_z+\omega q_z\Pi_{\parallel})^2}.\nonumber\\
\end{eqnarray}
Here, in $A_{i}$ only $A_{1}$ term contribute to the diffusion coefficients and can be written as
\begin{equation}
A_{1}=\mathcal{D}^{01}\tilde{\Delta}^{11}+\mathcal{D}^{02}\tilde{\Delta}^{12}+\mathcal{D}^{03}\tilde{\Delta}^{13}.
\label{a11}
\end{equation} 
The explicit terms in Eq.~(\ref{a11}) are given as
\begin{eqnarray}
\mathcal{D}^{01}\tilde{\Delta}^{11}&=&\frac{(D^{-1})^{01}j^{0}-(D^{-1})^{00}j^{1}}{(D^{-1})^{00}(D^{-1})^{11}-(D^{-1})^{10}(D^{-1})^{01}}\nonumber\\
&=&\frac{(v(q^2+q_z^2\Pi_{\parallel})-\omega q_x)2\pi\delta(\omega-\textbf{v}\cdot\textbf{q})}{(Q^2+q_x^2)(q^2+q_z^2\Pi_{\parallel})-\omega^2q_x^2}
\end{eqnarray}
\begin{eqnarray}
\mathcal{D}^{02}\tilde{\Delta}^{12}&=&\frac{(D^{-1})^{02}j^{0}-(D^{-1})^{00}j^{2}}{(D^{-1})^{00}(D^{-1})^{12}-(D^{-1})^{10}(D^{-1})^{02}}\nonumber\\
&=&\frac{-\omega 2\pi\delta(\omega-\textbf{v}\cdot\textbf{q})}{q_x(q^2+q_z^2\Pi_{\parallel})-\omega^2q_x}
\end{eqnarray}
\begin{eqnarray}
\mathcal{D}^{03}\tilde{\Delta}^{13}&=&\frac{(D^{-1})^{03}j^{0}-(D^{-1})^{00}j^{3}}{(D^{-1})^{00}(D^{-1})^{13}-(D^{-1})^{10}(D^{-1})^{03}}\nonumber\\
&=&\frac{(-\omega-\omega \Pi_{\parallel})2\pi\delta(\omega-\textbf{v}\cdot\textbf{q})}{q_x(q^2+q_z^2\Pi_{\parallel})-\omega^2q_x(1+\Pi_{\parallel})}.
\end{eqnarray}
The imaginary parts in the functions $A_{0}$ and $A_{i}$ can be obtained by splitting $\Pi_{\parallel}$ into its real and imaginary parts. 
 
%%%%%%%%%%%%%%%%%%%%%%%%%%%%%%%%%%%%%%%%%%%%%%%%%%%%%%%%%%%%%%%%%%


\begin{thebibliography}{99}

\bibitem{Kharzeev:NPA2008}
D. E. Khaezeev, L. D. McLerran and H. J. Warringa,
Nucl. Phys. A {\bf 803}, 227 (2008).

\bibitem{Skokov::JMPA2009}
V. Skokov, A. Y. Illarionov, and V. Toneev,
Int. J. Mod. Phys. A {\bf 24}, 5925 (2009).

\bibitem{Voronyuk:PRC2011}
V. Voronyuk, V. Toneev, W. Cassing, E. L. Bratkovskaya {\it et al.}
Phys. Rev. C {\bf 83}, 054911 (2011).

\bibitem{Deng: PRC2012}
W. Deng, and X. Huang,
Phys. Rev. C {\bf 85}, 044907 (2012).

\bibitem{Zhong:AHEP2014}
Y. Zhong, C. Yang, X. Cai, and S. Feng,
Adv. in High Ene. Phys., Vol-2014, Article: 193039.

%\cite{Adam:2019wnk}
\bibitem{Adam:2019wnk} 
  J.~Adam {\it et al.} [STAR Collaboration],
  %``First Observation of the Directed Flow of $D^{0}$ and $\overline{D^0}$ in Au+Au Collisions at $\sqrt{s_{\rm NN}}$ = 200 GeV,''
  Phys.\ Rev.\ Lett.\  {\bf 123}, no. 16, 162301 (2019).

%\cite{Acharya:2019ijj}
\bibitem{Acharya:2019ijj} 
  S.~Acharya {\it et al.} [ALICE Collaboration],
  %``Probing the effects of strong electromagnetic fields with charge-dependent directed flow in Pb-Pb collisions at the LHC,''
  arXiv:1910.14406 [nucl-ex].
     
\bibitem{Tuchin:PRC2011}
K. Tuchin,
Phys. Rev. C {\bf 83}, 017901 (2011);
K. Tuchin,
Adv. in High Ene. Phys. 2013, 1 (2013).

\bibitem{Fukushima:PRD2015}
K. Fukushima,
Phys. Rev. D {\bf 92}, 054009 (2015).

\bibitem{Mamo:PRD2015}
K. Mamo, and H. Yee,
Phys. Rev. D {\bf 92}, 105005 (2015).

\bibitem{Tuchin:PRC882013}
K. Tuchin,
Phys. Rev. C {\bf 88}, 024911 (2013);
K. Tuchin,
Phys. Rev. C {\bf 93}, 014905 (2016).

\bibitem{McLerran:NPA9292014}
L. McLerran, and V. Skokov,
Nucl. Phys. A {\bf 929}, 184 (2014).

\bibitem{Tuchin:2013ie} 
K.~Tuchin,
%``Particle production in strong electromagnetic fields in relativistic heavy-ion collisions,''
Adv.\ High Energy Phys.\  {\bf 2013}, 490495 (2013)
%doi:10.1155/2013/490495
[arXiv:1301.0099 [hep-ph]].

\bibitem{Yin:2013kya} 
Y.~Yin,
%``Electrical conductivity of the quark-gluon plasma and soft photon spectrum in heavy-ion collisions,''
Phys.\ Rev.\ C {\bf 90}, no. 4, 044903 (2014)
%doi:10.1103/PhysRevC.90.044903
[arXiv:1312.4434 [nucl-th]].

%\cite{Karmakar:2019tdp}
\bibitem{Karmakar:2019tdp} 
  B.~Karmakar, R.~Ghosh, A.~Bandyopadhyay, N.~Haque and M.~G.~Mustafa,
  %``Anisotropic pressure of deconfined QCD matter in presence of strong magnetic field within one-loop approximation,''
  Phys.\ Rev.\ D {\bf 99}, no. 9, 094002 (2019);
  B.~Karmakar, N.~Haque and M.~G.~Mustafa,
  %``Second-order quark number susceptibility of deconfined QCD matter in presence of magnetic field,''
  arXiv:2003.11247 [hep-ph].
       
    %\cite{Dash:2020vxk}
\bibitem{Dash:2020vxk} 
  A.~Dash, S.~Samanta, J.~Dey, U.~Gangopadhyaya, S.~Ghosh and V.~Roy,
  %``Anisotropic transport properties of Hadron Resonance Gas in magnetic field,''
  arXiv:2002.08781 [nucl-th]; 
  J.~Dey, S.~Satapathy, A.~Mishra, S.~Paul and S.~Ghosh,
  %``From Non-interacting to Interacting Picture of Quark Gluon Plasma in presence of magnetic field and its fluid property,''
  arXiv:1908.04335 [hep-ph].
    
%\cite{Hattori:2017qih}
\bibitem{Hattori:2017qih} 
  K.~Hattori, X.~G.~Huang, D.~H.~Rischke and D.~Satow,
  %``Bulk Viscosity of Quark-Gluon Plasma in Strong Magnetic Fields,''
  Phys.\ Rev.\ D {\bf 96}, no. 9, 094009 (2017);
  K.~Hattori and D.~Satow,
  %``Electrical Conductivity of Quark-Gluon Plasma in Strong Magnetic Fields,''
  Phys.\ Rev.\ D {\bf 94}, no. 11, 114032 (2016).   
  
\bibitem{Kurian:2017yxj} 
  M.~Kurian and V.~Chandra,
  Phys.\ Rev.\ D {\bf 96}, no. 11, 114026 (2017).

\bibitem{fukushima:PRD782008}
K. Fukushima, D. E. Kharzeev, and H. Warringa,
Phys. Rev. D {\bf 78}, 074033 (2008).

\bibitem{Kharzeev::PRD832011}
D. E. Kharzeev, and H. Yee,
Phys. Rev. D {\bf 83}, 085007 (2011).

\bibitem{Newman:JHEP012006}
G. M. Newman,
J. High Energy Phys. {\bf 158}, 01 (2006).

\bibitem{Burnier:PRL1072011}
Y. Burnier, D. E. Kharzeev, J. Liao, and H. Yee,
Phys. Rev. Lett. {\bf 107}, 052303 (2011).

\bibitem{Gorbar:PRD832011}
E. Gorbar, V. Miransky, and I. Shovkovy,
Phys. Rev. D {\bf 83}, 085003 (2011).

 %\cite{Gusynin:1995nb}
\bibitem{Gusynin:1995nb} 
  V.~P.~Gusynin, V.~A.~Miransky and I.~A.~Shovkovy,
  %``Dimensional reduction and catalysis of dynamical symmetry breaking by a magnetic field,''
  Nucl.\ Phys.\ B {\bf 462}, 249 (1996).

\bibitem{Feng:PRD962017}
B. Feng,
Phys. Rev. D {\bf 96}, 036009 (2017).

%\cite{Fukushima:2017lvb}
\bibitem{Fukushima:2017lvb} 
  K.~Fukushima and Y.~Hidaka,
  %``Electric conductivity of hot and dense quark matter in a magnetic field with Landau level resummation via kinetic equations,''
  Phys.\ Rev.\ Lett.\  {\bf 120}, no. 16, 162301 (2018).

%\cite{Kurian:2017yxj}
\bibitem{Kurian:2018qwb} 
M.~Kurian, S.~Mitra, S.~Ghosh and V.~Chandra,
  %``Transport coefficients of hot magnetized QCD matter beyond the lowest Landau level approximation,''
  Eur.\ Phys.\ J.\ C {\bf 79}, no. 2, 134 (2019);
   M.~Kurian and V.~Chandra,
  %``Effective description of hot QCD medium in strong magnetic field and longitudinal conductivity,''
  Phys.\ Rev.\ D {\bf 96}, no. 11, 114026 (2017).
    
\bibitem{Bandyopadhyay:PRD2016}
A. Bandyopadhyay, C. A. Islam, and M. G. Mustafa,
Phys. Rev. D {94}, 114034 (2016);
  A.~Das, N.~Haque, M.~G.~Mustafa and P.~K.~Roy,
  %``Hard dilepton production from a weakly magnetized hot QCD medium,''
  Phys.\ Rev.\ D {\bf 99}, no. 9, 094022 (2019).

\bibitem{Tuchin:PRC832011}
K. Tuchin,
Phys. Rev. C {\bf 83}, 017901 (2011).
  
\bibitem{Ghosh:2018xhh} 
  S.~Ghosh and V.~Chandra,
  %``Electromagnetic spectral function and dilepton rate in a hot magnetized QCD medium,''
  Phys.\ Rev.\ D {\bf 98}, no. 7, 076006 (2018).

\bibitem{CS:2018mag} 
A.~Mishra, A.~Jahan CS, S.~Kesarwani, H.~Raval, S.~Kumar and J.~Meena,
%``Charmonium decay widths in magnetized matter,''
Eur.\ Phys.\ J.\ A {\bf 55}, no. 6, 99 (2019)
%doi:10.1140/epja/i2019-12778-2
[arXiv:1812.07397 [nucl-th]].

\bibitem{Reddy:2017pqp} 
S.~Reddy P., A.~Jahan C. S., N.~Dhale, A.~Mishra and J.~Schaffner-Bielich,
%``D mesons in strongly magnetized asymmetric nuclear matter,''
Phys.\ Rev.\ C {\bf 97}, no. 6, 065208 (2018)
%doi:10.1103/PhysRevC.97.065208
[arXiv:1712.07997 [nucl-th]].

\bibitem{Hasan:EPJC2017}
M. Hasan, B. Chatterjee, and B. K. Patra,
Eur. Phys. J. C {\bf 77}, 767 (2017).

\bibitem{Singh:PRD972018}
B. Singh, L. Thakur, and H. Mishra,
Phys. Rev. D {\bf 97}, 096011 (2018).

%\cite{Finazzo:2016mhm}
\bibitem{Finazzo:2016mhm} 
  S.~I.~Finazzo, R.~Critelli, R.~Rougemont and J.~Noronha,
  %``Momentum transport in strongly coupled anisotropic plasmas in the presence of strong magnetic fields,''
  Phys.\ Rev.\ D {\bf 94}, no. 5, 054020 (2016).

\bibitem{Fukushima:2015wck} 
 K.~Fukushima, K.~Hattori, H.~U.~Yee and Y.~Yin,
  %``Heavy Quark Diffusion in Strong Magnetic Fields at Weak Coupling and Implications for Elliptic Flow,''
Phys.\ Rev.\ D {\bf 93}, no. 7, 074028 (2016).

\bibitem{Kurian:2019nna} 
M.~Kurian, S.~K.~Das and V.~Chandra,
  %``Heavy quark dynamics in a hot magnetized QCD medium,''
Phys.\ Rev.\ D {\bf 100}, no. 7, 074003 (2019);  
M.~Kurian, V.~Chandra and S.~K.~Das,
  %``Impact of longitudinal bulk viscous effects to heavy quark transport in a strongly magnetized hot QCD medium,''
  arXiv:2002.03325 [nucl-th].

\bibitem{Svetitsky:1987gq} 
B.~Svetitsky,
  %``Diffusion of charmed quarks in the quark-gluon plasma,''
Phys.\ Rev.\ D {\bf 37}, 2484 (1988).
  
\bibitem{GolamMustafa:1997id}
M.~Golam Mustafa, D.~Pal and D.~Kumar Srivastava,
  %``Propagation of charm quarks in equilibrating quark - gluon plasma,''
Phys.\ Rev.\ C {\bf 57}, 889 (1998)
Erratum: [Phys.\ Rev.\ C {\bf 57}, 3499 (1998)]. 
  
\bibitem{Moore:2004tg} 
G.~D.~Moore and D.~Teaney,
  %``How much do heavy quarks thermalize in a heavy ion collision?,''
Phys.\ Rev.\ C {\bf 71}, 064904 (2005).
  
 %\cite{CaronHuot:2007gq}
\bibitem{CaronHuot:2007gq} 
S.~Caron-Huot and G.~D.~Moore,
  %``Heavy quark diffusion in perturbative QCD at next-to-leading order,''
Phys.\ Rev.\ Lett.\  {\bf 100}, 052301 (2008).

  %\cite{vanHees:2005wb}
\bibitem{vanHees:2005wb} 
H.~van Hees, V.~Greco and R.~Rapp,
  %``Heavy-quark probes of the quark-gluon plasma at RHIC,''
Phys.\ Rev.\ C {\bf 73}, 034913 (2006);
H.~van Hees, M.~Mannarelli, V.~Greco and R.~Rapp,
  %``Nonperturbative heavy-quark diffusion in the quark-gluon plasma,''
Phys.\ Rev.\ Lett.\  {\bf 100}, 192301 (2008).

 \bibitem{Das:2013kea}  
  S.~K.~Das, F.~Scardina, S.~Plumari and V.~Greco,
  %``Heavy-flavor in-medium momentum evolution: Langevin versus Boltzmann approach,''
  Phys.\ Rev.\ C {\bf 90}, 044901 (2014).

%\cite{Singh:2018wps}
\bibitem{Singh:2018wps} 
  B.~Singh, A.~Abhishek, S.~K.~Das and H.~Mishra,
  %``Heavy quark diffusion in a Polyakov loop plasma,''
  Phys.\ Rev.\ D {\bf 100}, no. 11, 114019 (2019);
  B.~Singh and H.~Mishra,
  %``Heavy quark transport in a viscous semi QGP,''
  arXiv:1911.06764 [hep-ph].
  %%CITATION = ARXIV:1911.06764;%%
  
  %\cite{Das:2012ck}
\bibitem{Das:2012ck} 
  S.~K.~Das, V.~Chandra and J.~e.~Alam,
  %``Heavy-quark transport coefficients in a hot viscous quark–gluon plasma medium,''
  J.\ Phys.\ G {\bf 41}, 015102 (2013); 
V.~Chandra and S.~K.~Das,
%``Impact of momentum-space anisotropy on heavy quark dynamics in a QGP medium,''
 Phys. Rev. D \textbf{93}, no.9, 094036 (2016); 
  S.~K.~Das, J.~e.~Alam and P.~Mohanty,
  %``Probing quark gluon plasma properties by heavy flavours,''
  Phys.\ Rev.\ C {\bf 80}, 054916 (2009).

\bibitem{Cao:2018ews}
  S.~Cao {\it et al.},
  %``Toward the determination of heavy-quark transport coefficients in quark-gluon plasma,''
  Phys.\ Rev.\ C {\bf 99}, no. 5, 054907 (2019).
  
  \bibitem{Rapp:2018qla}
  R.~Rapp {\it et al.},
  %``Extraction of Heavy-Flavor Transport Coefficients in QCD Matter,''
  Nucl.\ Phys.\ A {\bf 979}, 21 (2018). 
  
  \bibitem{Giataganas:2013zaa} 
D.~Giataganas and H.~Soltanpanahi,
  %``Heavy Quark Diffusion in Strongly Coupled Anisotropic Plasmas,''
  JHEP {\bf 1406}, 047 (2014);
D.~Giataganas and H.~Soltanpanahi,
  %``Universal Properties of the Langevin Diffusion Coefficients,''
Phys.\ Rev.\ D {\bf 89}, no. 2, 026011 (2014).

\bibitem{Alberico:2013bza} 
  W.~M.~Alberico, A.~Beraudo, A.~De Pace, A.~Molinari, M.~Monteno, M.~Nardi, F.~Prino and M.~Sitta,
  %``Heavy flavors in $AA$ collisions: production, transport and final spectra,''
  Eur.\ Phys.\ J.\ C {\bf 73}, 2481 (2013).
  
  \bibitem{Aarts:2016hap}
  G.~Aarts {\it et al.},
  %``Heavy-flavor production and medium properties in high-energy nuclear collisions - What next?,''
  Eur.\ Phys.\ J.\ A {\bf 53}, no. 5, 93 (2017)
 
\bibitem{Dong:2019unq}
  X.~Dong and V.~Greco,
  %``Heavy quark production and properties of Quark–Gluon Plasma,''
  Prog.\ Part.\ Nucl.\ Phys.\  {\bf 104}, 97 (2019).
  
\bibitem{Cao:2013ita} 
  S.~Cao, G.~Y.~Qin and S.~A.~Bass,
  %``Heavy-quark dynamics and hadronization in ultrarelativistic heavy-ion collisions: Collisional versus radiative energy loss,''
  Phys.\ Rev.\ C {\bf 88}, 044907 (2013).
  
\bibitem{Song:2015sfa} 
  T.~Song, H.~Berrehrah, D.~Cabrera, J.~M.~Torres-Rincon, L.~Tolos, W.~Cassing and E.~Bratkovskaya,
  %``Tomography of the Quark-Gluon-Plasma by Charm Quarks,''
  Phys.\ Rev.\ C {\bf 92}, no. 1, 014910 (2015).
  
\bibitem{Scardina:2017ipo} 
  F.~Scardina, S.~K.~Das, V.~Minissale, S.~Plumari and V.~Greco,
  Phys.\ Rev.\ C {\bf 96}, no. 4, 044905 (2017);
  S.~Plumari, V.~Minissale, S.~K.~Das, G.~Coci and V.~Greco,
  Eur.\ Phys.\ J.\ C {\bf 78}, no. 4, 348 (2018).
    
\bibitem{Adare:2006nq} 
  A.~Adare {\it et al.} [PHENIX Collaboration],
  %``Energy Loss and Flow of Heavy Quarks in Au+Au Collisions at s(NN)**(1/2) = 200-GeV,''
  Phys.\ Rev.\ Lett.\  {\bf 98}, 172301 (2007);
  S.~S.~Adler {\it et al.} [PHENIX Collaboration],
  %``Nuclear modification of electron spectra and implications for heavy quark energy loss in Au+Au collisions at s(NN)**(1/2) - 200-GeV,''
  Phys.\ Rev.\ Lett.\  {\bf 96}, 032301 (2006).  
   
 \bibitem{Andronic:2015wma}
  A.~Andronic {\it et al.},
  %``Heavy-flavour and quarkonium production in the LHC era: from proton–proton to heavy-ion collisions,''
  Eur.\ Phys.\ J.\ C {\bf 76}, no. 3, 107 (2016).
  
  \bibitem{Singh:2020fsj}
B.~Singh, S.~Mazumder and H.~Mishra,
%``HQ Collisional energy loss in a magnetized medium,''
[arXiv:2002.04922 [hep-ph]].
  
% \bibitem{Romatschke:2007PRC}
%P. Romatschke,
%  %Momentum broadening in an anisotropic plasma
%Phys.\ Rev.\ C {\bf 75}, 014901 (2007).  

%\cite{Romatschke:2006bb}
\bibitem{Romatschke:2006bb}
P.~Romatschke,
%``Momentum broadening in an anisotropic plasma,''
Phys. Rev. C \textbf{75}, 014901 (2007)
%doi:10.1103/PhysRevC.75.014901
[arXiv:hep-ph/0607327 [hep-ph]].
%108 citations counted in INSPIRE as of 18 Aug 2020

\bibitem{Romatschke:thesis}
P. Roamtschke, arXiv:hep-ph/0312152.

\bibitem{Mamo:2016prd}
S.~Li, K.~A.~Mamo and H.~Yee,
Phys. Rev. D \textbf{94}, 085016 (2016).

\bibitem{Das:2016cwd} 
  S.~K.~Das, S.~Plumari, S.~Chatterjee, J.~Alam, F.~Scardina and V.~Greco,
  %``Directed Flow of Charm Quarks as a Witness of the Initial Strong Magnetic Field in Ultra-Relativistic Heavy Ion Collisions
  Phys.\ Lett.\ B {\bf 768}, 260 (2017).
    
\bibitem{Chatterjee:2018lsx} 
  S.~Chatterjee and P.~Bozek,
  %``Interplay of drag by hot matter and electromagnetic force on the directed flow of heavy quarks,''
  Phys.\ Lett.\ B {\bf 798}, 134955 (2019).

\bibitem{Hattori:2012ny}
K.~Hattori and K.~Itakura,
%``Vacuum birefringence in strong magnetic fields: (II) Complex refractive index from the lowest Landau level,''
Annals Phys.\  \textbf{334} (2013), 58-82.








 
  
\end{thebibliography}
\end{document}